\documentclass[10pt,twoside]{article}
\usepackage{hyperref}
\usepackage{amssymb}
\usepackage[mathscr]{eucal}
\usepackage{eufrak}
\usepackage{amsmath}
\usepackage{mathrsfs}

\def\intl#1{\int\limits_{#1}}
\def\intll#1#2{\int\limits_{#1}^{#2}}
\def\dm{|\hskip-0.05cm|}
\def\OO{\Omega}
\def\displ{\displaystyle}
\def\VSE{\vspace{6pt}\\&\displ }
\def\VS{\vspace{6pt}\\\displ }
\def\rf#1{{\rm(\ref{#1})}}
\def\chiu{\hfill$\displaystyle\vspace{4pt}
\underset\Box\null$\par}
\def\Pr{{\bf Proof. }}

\def\R{\Bbb R}
\def\N{\Bbb N}
\def\o{\"{o}}
\def\à{\`{a}}
\def\è{\`{e}}
\def\ì{\`{i}}
\def\ù{\`{u}}
\def\ò{\`{o}}
\def\é{\'{e}}

\def\dy{\displaystyle}
\def\vep{\varepsilon}

\def\be{\begin{equation}}
\def\ba{\begin{array}}
\def\ea{\end{array}}
\def\ee{\end{equation}}
\def\vs1{\vspace{1ex}}
\def\vp{\varphi}
\def\ov{\overline}

\font\sc=cmcsc10
 \title{On the spatial asymptotic decay of a suitable weak solution to the Navier-Stokes Cauchy problem}
\author{\sc F. Crispo and P. Maremonti
\thanks{
Dipartimento di Matematica e Fisica, Seconda
Universit\`{a} degli Studi di
 Napoli, via Vivaldi 43, 81100 Caserta,
 Italy.
francesca.crispo@unina2.it;
paolo.maremonti@unina2.it}}
\date{}
\begin{document}

\maketitle \noindent{\bf Abstract} - {\small   We prove space-time decay estimates of suitable weak solutions to the Navier-Stokes Cauchy problem, corresponding to a given asymptotic behavior of the initial data of the same order of decay. We use two main tools. The first is a result obtained in  \cite{CMRWS} on the behavior of the solution in a neighborhood of $t=0$ in the $L^\infty_{loc}$-norm,  which enables us to furnish a representation formula for a suitable weak solution. The second is the asymptotic behavior of $\dm u(t)\dm_{L^2(\R^3\setminus B_R)}$ for $R\to\infty$. Following  a Leray's point of view, roughly speaking our result proves that a possible space-time turbulence does not perturb the asymptotic spatial behavior of the initial data of a suitable weak solution. }
 \par\noindent
 \vskip -0.7true cm\noindent
\newcommand{\red}{\protect\bf}
\renewcommand\refname{\centerline
{\red {\normalsize \bf References}}}
\newtheorem{ass}
{\bf Assumption}[section]
\newtheorem{defi}
{\bf Definition}[section]
\newtheorem{tho}
{\bf Theorem}[section]
\newtheorem{rem}
{\sc Remark}[section]
\newtheorem{lemma}
{\bf Lemma}[section]
\newtheorem{coro}
{\bf Corollary}[section]
\newtheorem{prop}
{\bf Proposition}[section]
\renewcommand{\theequation}{\thesection.\arabic{equation}}
\setcounter{section}{0}
\numberwithin{equation}{section}
\section{\large Introduction}
In this paper, we study the initial   value problem:
\be\label{NS}\ba{l}v_t+v\cdot
\nabla v+\nabla\pi_v=\Delta
v,\;\nabla\cdot
v=0,\mbox{ in }(0,T)\times\R^3,\\
 v(0,x)=v_\circ(x)\mbox{
on }\{0\}\times\R^3.\ea\ee In system
\rf{NS} $v$ is the kinetic field,
$\pi_v$ is the pressure field,
 $v_t\!:=\!
\frac\partial{\partial t}v$  and
 $v\cdot\nabla v\!:=\!
v_k\frac\partial{\partial x_k}v$. 
 For brevity, we assume zero
body force.  
  We set $J^q(\R^3)\!:=$completion of
$\mathscr C_0(\R^3)$ with respect to the $L^q$-norm, $q\in(1,\infty)$. The symbol
$\mathscr C_0(\R^3)$ denotes the subset of
$C_0^\infty(\R^3)$ whose elements are divergence
free. By $P_q$ (the index $q$ is omitted when there is no danger of confusion) we mean the projector from $L^q$  into $J^q$. For properties and details on these spaces see for instance \cite{G}. Moreover, we set $J^{1,q}(\R^3)\!:=$completion of $\mathscr C_0(\R^3)$ with respect to the $W^{1,q}$-norm. For a nonnegative integer $m$, if $X$ is a Banach space, the symbols $C^m(a,b;X)$ and $L^p(a,b;X)$ mean the spaces of functions defined in $(a,b)\subseteq \R$ with value in the Banach space $X$, that are $m$-times continuous differentiable in $[a,b]$ and $L^p$-integrable on $(a,b)$, respectively. \par We use the same symbol to denote  vector or scalar functions and function spaces.   We set
$(u,g):=\int\limits_{\!\!\R^3} u\cdot
gdx\,.$
\begin{defi}\label{WS}{\sl  A pair $(v,\pi_v)$, such that $v:(0,T)\times\R^3 \to\R^3$ and $\pi_v:(0,T)\times\R^3 \to \R$, is said a weak solution to problem {\rm\rf{NS}} if
\begin{itemize}\item [i)] for all $T>0$,
 $v\in  L^2(0,T; J^{1,2}
(\R^3 ))$ and $\pi_v\in L^\frac53((0,T)\times\R^3)$, $$\dm v(t)\dm_2^2+2\intll st \dm\nabla v(\tau)\dm_2^2d\tau\leq \dm v(s)\dm_2^2,\;\forall t\geq s, \mbox{ for } s=0 \mbox{ and a.e. in } s\geq 0,$$\item
[ii)] $\displ\lim_{t\to0}\dm
v(t)-v_\circ\dm_2=0$,\item[iii)]
for all $t,s\in(0,T)$,  the pair $(v,\pi_v)$
satisfies the  equation:
\newline{$\displ\intll
st\Big[(v,\varphi_\tau)-(\nabla
v,\nabla
\varphi)+(v\cdot\nabla\varphi,v)+(\pi_v,\nabla
\cdot\varphi)\Big]d\tau+(v(s),\varphi
(s))=(v(t),\varphi(t))$,}
\newline\centerline{
 for all $\varphi\in C^1_0([0,T)\times\R^3 )$.}
\end{itemize}}\end{defi} 
As it is known, the regularity and uniqueness of a weak solution are still  open problems. However, in order to improve  the results of regularity, in the fundamental paper \cite{CKN},   Caffarelli, Kohn and Nirenberg introduce the notion of suitable weak solution: 
\begin{defi}\label{SWS}{\sl A pair $(v,\pi_v)$ is said a suitable weak solution if it is a weak solution in the sense of the Definition\,\ref{WS} and, moreover,
\be\label{SEI}\ba{l}\displ\intl{\R^3}|v(t)|^2\phi(t)dx+2\intll st\intl
{\R^3}|\nabla v|^2\phi dxd\tau\leq \intl{\R^3}|v(s)|^2\phi(s)dx\VS\hskip 2,5cm+\intll st\intl{\R^3}|v|^2(\phi_\tau+\Delta\phi)dxd\tau+\intll st\intl{\R^3}(|v|^2+2\pi_v)
v\cdot\nabla\phi dxd\tau,\ea\ee for all $t\geq s$, for $s=0$ and a.e. in $s\geq 0$, and for all 
nonnegative $\phi\in C_0^\infty(\R\times\R^3)$.}\end{defi}
In \cite{CKN} and \cite{scheffer1} the following existence result is proved:
\begin{tho}\label{EXCKN}{\sl For all $v_\circ\in J^2(\OO)$ there exists a suitable weak solution.}\end{tho}
Further in the same paper \cite{CKN} it is proved that  in the set of suitable weak solutions a partial regularity result holds. Among others things, in \cite{CKN} the authors  prove that under the further assumption $\nabla v_\circ\in L^2(\R^3\setminus B_{R_0})$ a suitable weak solution is regular in a neighborhood of {\it ``infinity''}, that is for $|x|>M_0R_0\,, M_0>1$. Roughly speaking, this result is the analogous of the one related to the {\it structure theorem} by Leray, with the difference that Leray's result is on regularity of a weak solution in neighborhood of $t=+\infty$. Still in the light of the analogy, roughly speaking, following Leray, we say that   the possible {\it turbulence} of a weak solution  not only appears in a finite time but also in a bounded region of the space, whose parabolic one-dimensional Hausdorff measure is null (this last is true as soon as a suitable weak solution exists). Actually, the smallness of the data for large  $|x|$, although given by means of integrability conditions outside a ball, preserves the  regularity of the weak solution (as for {\it ``small data''}).\par In the wake of the previous results, in \cite{CMRWS} we have proved a result concerning the behavior in time of the $L^\infty_{loc}$ norm of the solution in a neighborhood of $t=0$ for suitable weak solutions, corresponding to a suitably small data (see Theorem \ref{CT} below).
\par The aim of this note goes in the direction of the last claims. We prove that, not only the possible {\it turbulence} does not perturb 
the regularity of a weak solution in a neighborhood of $t=+\infty$, but, if  an asymptotic spatial behavior  of the initial data $v_\circ$ is given, then the same behavior holds  for a suitable weak solution for all $t>0$. As far as we know, such a property to date was ensured only for small data (cf. \cite{CMSB}, \cite{K}).\par
The theorem we are going to state is the main result of the paper:
\begin{tho}\label{CR}{\sl Let $v_\circ\in J^2(\R^3)$ and, for some $\alpha\in[1,3)$ and 
$R_0>0$, let be $|v_\circ(x)|\leq V_\circ|x|^{-\alpha}$, for $|x|>R_0$. Let $(v,\pi_v)$ be a suitable weak solution  
to the Navier-Stokes Cauchy problem. Then, there exists a
 constant $M\geq1$ such that \be\label{CRI}|v(t,x)|\leq c(v_\circ)|x|^{-\alpha},\mbox{ for all }(t,x)\in (0,\infty)\times \R^3\setminus B_{MR_0},\ee
where $M$ is independent of $v_\circ$ and $c(v_0)$ depends on $V_\circ$ and $\dm v_\circ\dm _2$.}\end{tho}\begin{coro}\label{CCR}{\sl For a solution of Theorem\,\ref{CR}, for all $\beta\in[0,\alpha]$,  
we get:\be\label{CCRI}|v(t,x)|\leq c(v_\circ)|x|^{-\alpha+\beta}t^{-\frac\beta2},\mbox{ for all }(t,x)\in (0,\infty)\times \R^3\setminus B_{MR_0};\ee
and \be\label{CCRII}|v(t,x)|\leq c(v_\circ)|x|^{-\alpha+\beta}t^{-\frac\beta2},\mbox{ for all }(t,x)\in (T_0,\infty)\times \R^3,\ee where $T_0\leq c\dm v_\circ\dm_2^4$.}\end{coro}
 \par Here we state Theorem\,\ref{CR} for solutions to the Navier-Stokes Cauchy problem just for the sake of brevity. 
 The result of Theorem\,\ref{CR} can be seen as a continuous dependence of the null solution. In this sense the theorem is a continuation of the one proved in \cite{CMRWS}, that we employ here for regularity, see Theorem\,\ref{CT} below.\par
In a forthcoming paper, the same result will be proved for a three dimensional exterior domain $\OO$ and for weak solutions with initial data in $J^3(\OO)$, hence not necessarily with finite energy! This assumption seems to be more coherent with the assumption $|v_\circ(x)|\leq V_\circ |x|^{-\alpha},\,\alpha\in[1,3)$.
\par As far as we know, the technique employed to prove Theorem\,\ref{CR} is original in the framework of the ones employed to prove a spatial asymptotic behavior of a solution $v(t,x)$, more in general, to a parabolic equation. Indeed, it essentially  follows from two properties. The former, well known, concerns the spatial behavior of the solution $w(t,x)$ to the heat equation. The latter is connected with the asymptotic behavior of the functional:
\be\label{SEIS}\intl{|y|>c|x|}|u(t,y)|^2dy\leq c(t)|x|^{-1},\ee where $u=v-w$, which we think to be an original tool for the spatial behavior. Estimate \rf{SEIS} comes from   estimate \rf{SEI} for a suitable $\phi(x)$, written for the difference $u$. For the special $\phi$, we like to call the above functional as Leray's generalized energy inequality. It was used by Leray in \cite{L} for a compactness property.   \par We conclude the introduction by quoting paper \cite{PD}, where a similar result is given.
More precisely, for the Navier-Stokes initial boundary value problem in an exterior domain $\OO\subseteq\R^3$ it is proved:
\par{\it Assume that the initial data $v_0\in J^2(\OO)\cap J^\frac54(\OO)$ with $|v_0(x)| \leq V_0|x|^{-\alpha}
$, for some $\alpha\in [\frac76,3)$ and $|x|>R_0>0$.
Assume that, $\ov\gamma\in(0,\frac14]$, $v_0\in D(A^{\frac34+\ov\gamma}_2)\cap J^\frac98(\OO)$\, . Then, there are suitable constants $c_0$ and $\ov R_0$ such that
$$|v(t,x)|\leq c_0|x|^{-\min\{\alpha,\frac{13}5\}},\mbox{ for all }(t,x)\in (0,\infty)\times \R^3\setminus B_{\ov R_0}.$$}
\par\noindent Here, $A_2:=P_2\Delta$ is the Stokes operator, and  $D(A^s_2)$ is the domain of definition of the fractional power $s$ of $A_2$.
\par We point out that key ideas, the technique and the proofs in  \cite{PD} and in the present paper are completely different.
\vskip0.05cm\par The plan of the paper is the following. In order to perform pointwise estimates, in sec.\,2 we establish results of partial regularity based on the results of \cite{CMRWS}. In sec.\,3 we recall classical results concerning the solutions to the Stokes Cauchy problem. In sec.\,4 we prove estimate \rf{SEIS} which is strategic for our aims. In sec.\,5 we give the representation formula of the solutions to the Navier-Stokes Cauchy problem that we employ in sec.\,6 and sec.\,7 to prove our results. \section{\large Preliminary results on partial regularity of a suitable weak solution}
Throughout the paper, where it is appropriate, we give an explicit dependence of the constants from the $L^2$-norm of the data. In the other cases, the dependence will be referred to simply by $c(v_\circ)$. \begin{lemma}\label{SD}{\sl In the hypotheses of Theorem\,\ref{CR} 
\be\label{SDI}\ba{l} \mbox{for all }\vep>0,\;\displ\intl{\R^3}\frac{|v_\circ(y)|}{|x-y|}\!{{\null^2}\atop\null}dy\leq \frac c{\vep|x|}{\dm v_\circ\dm_2^2}+\frac c{|x|}\null_\frac12V_\circ\dm v_\circ\dm_2,\;|x|\!>\!\frac{R_0}{1-\!\vep}, \ea \ee with $c$ independent of $x$ and $v_\circ$}\end{lemma}
\Pr We start by proving \rf{SDI} for $\alpha=1$. Then, {\it a fortiori}, it holds for $\alpha\in(1,3)$. Given $\vep>0$ and $x\in\R^3$, by virtue of our assumption,  we easily deduce that 
\be\label{SDII}\ba{ll}\displ\intl{\R^3}\frac{|v_\circ|^2}{|x-y|}dy\hskip-0.2cm&\displ\leq\hskip-0.2cm
\intl{|y|<(1-\vep)|x|}\hskip-0.2cm\frac{|v_\circ|^2}{|x-y|}dy+cV_\circ\hskip-1cm\intl{\hskip0.5cm(1-\vep)|x|<|y|}\hskip-0.2cm\frac{|v_\circ|}{|x-y|(1+|y|)}dy \VSE
\leq \frac{\dm v_\circ\dm_2 }{\vep|x|}+ c\Big[\hskip-0.3cm\intl{\hskip0.3cm(1-\vep)|x|<|y|}\hskip-0.5cm\frac{(1+|y|)^{-2}}{|x-y|^2}dy\Big]^\frac12V_\circ\dm v_\circ\dm_2
,\ea\ee which implies the thesis.    \chiu
\par Let $ x_0\in \R^3$ and $R_0>0$.  Let  $v_\circ\in J^2(\R^3 )$. We set  
$${\mathscr E}_0(x_0,R_0):=\mbox{ess}\hskip-0.1cm\sup_{\hskip-0.4cmB(x_0,R_0)}\dm
v_\circ\dm_{w(x)}:=\dm\big(\intl{\R^3}\frac{|v_\circ(y)|^2}{|x-y|}
dy\big)^\frac12\dm_{L^\infty(B(x_0,R_0))}
.$$
\begin{tho}\label{CT}{\sl 
Let $(v,\pi_v)$ be a suitable weak solution corresponding to $v_\circ\in J^2(\R^3)$. There exist absolute constants $\vep_1$, $C_1$ and $C_2$ such that, if 
\be\label{th}\;
C_1\mathscr E_0( x_0,R_0)<1\mbox{ and }C_2(\mathscr E_0^3+\mathscr E_0^\frac52)\leq\vep_1,\ee
then 
\be\label{thh}|v(t,x)|\leq
c(\mathscr E_0^3+\mathscr E_0^\frac52)^\frac13t^{-\frac12},\ee provided that $(t,x)$ is a Lebesgue point with $\dm v_\circ\dm_{w(x)}<\infty$ and $x\in B(x_0,R_0)$.}
\end{tho}\Pr See \cite{CMRWS}, Theorem\,1.1.\chiu 
\par We complete the results concerning a suitable weak solution with the following lemma on the regularity and the asymptotic behavior of the solutions:
\begin{lemma}\label{DS}{\sl If $v_\circ\in J^2(\R^3)\cap J^p(\R^3),p\in(1,2]$, then there exists a $T_0\leq c\dm v_\circ\dm_2^4 $ such that \be\label{DE}\ba{ll}&\dm v(t)\dm_2=\left\{\ba{ll}o(1)&\mbox{if }p=2,\\ c(v_\circ)t^{-\frac 32\left(\frac1p-\frac12\right)},\;t>0&\mbox{if }p\in(1,2).\ea\right.
\VSE \dm v(t)\dm_\infty\leq c\dm v_\circ\dm_2t^{-\frac34},\;t>T_0.\ea\ee}\end{lemma}
\Pr Estimates \rf{DE}$_{1,2}$ can be found in \cite{MCMP}, instead \rf{DE}$_3$ is well known (see Leray \cite{L}).
 \begin{lemma}\label{CCT}{\sl In the hypotheses of Theorem\,\ref{CR} there exists a constant  $M_0\geq 1$ such that \be\label{CCTI}|v(t,x)|\leq 
c(\mathscr E_0^3+\mathscr E_0^\frac52)^\frac13t^{-\frac12},\mbox{ a.e. in }t>0\mbox{ and }|x|>M_0R_0,\ee provided that $(t,x)$ is a Lebesgue point.}
\end{lemma}
\Pr It is enough to verify that the hypotheses of Theorem\,\ref{CT} are satisfied. To this end, we  employ Lemma\,\ref{SD} which ensures the existence of $\ov R(C_1, C_2,\vep_1)$ such that, for $|x|>\ov R$,  
$[\intl{\R^3}\frac{|v_\circ(x)|}{|x-y|}\!{{\null^2}\atop\null}dy]^\frac12$ satisfies \rf{th}. Setting $\ov R:=M_0R_0$, we have proved the lemma.\chiu As a consequence of the above lemmas on the $L^\infty$-norm of a suitable weak solution we can claim
\begin{coro}\label{LIO}{\sl In the hypotheses of Theorem\,\ref{CR}, we get
\be\label{LIOI}\dm v(t)\dm_{L^\infty(|x|>M_0R_0)}\leq c(v_\circ,T_0)t^{-\frac34},\;t>0.\ee}\end{coro}
\begin{lemma}\label{PWS}{\sl If $(v,\pi_v)$ is a suitable weak solution, then, the pressure field admits the representation formula:
\be\label{PWSI}
\pi_v(t,x)=-D_{x_i}D_{x_j}\intl{\R^3}\mathcal E(x-y)v^i(y)v^j(y)dy=:\mathbb E[v,v](t,x)\,,\ee a.e. in $(t,x)\in (0,\infty)\times\R^3$.}\end{lemma}\Pr See  \cite{CMRWS}, Lemma\,4.1.
\section{\large Some lemmas on the Stokes Cauchy problem}
We consider the Stokes Cauchy problem
\be\label{HSO}w_t-\Delta w=-\nabla \pi_w,\;\nabla \cdot w=0\mbox{ in }(0,T)\times \R^3,\;w(0,x)=v_\circ(x)\mbox{ on }\{0\}\times \R^3.\ee
We denote by 
\be\label{HS00}H_{ij}(s,z):=\delta_{ij}H(s,z):=\delta_{ij}(4\pi s)^{-\frac32}e^{-\frac{|z|^2}{4s}}\ee the general component of the heat kernel tensor, and set
$$\mathbb H[w(s)](t-s,x):=\intl{\R^3} H(t-s,x-y) w(s,y)dy\,.$$ 
Then, for the solution  of \eqref{HSO} we have $w(t,x):=\mathbb H[v_\circ](t,x)$. Moreover, we recall the estimate ($k$ nonnegative integer and $\beta$ multi-index):
\be\label{HSII}|D^k_sD^\beta_zH(s,z)|
\leq c(|z|+s^\frac12)^{-3-2k-|\beta|}\,.\ee 
\par  We are interested in the following result:
\begin{lemma}\label{HS}{\sl In the hypotheses of Theorem\,\ref{CR} on $v_\circ$,  \be\label{HS0}\ba{ll}
\mbox{for all } T>0,  w\in C(0,T; L^2(\R^3))\,,\VS 
\mbox{for all } \eta>0,k\geq0, \,|\beta|\geq0,D^k_tD^\beta_x w\in C(\eta,T; C_b(\R^3))\cap C(\eta,T;L^2(\R^3))\,,\VS \dm w(t)\dm+2\intll st\dm \nabla w(\tau)\dm^2d\tau=\dm w(s)\dm^2,\mbox{ for all }t\geq s\geq0.\ea\ee Moreover, there holds\be\label{HSI}
|w(t,x)|\leq c\min\{\dm v_\circ\dm_2,V_\circ\}\min\big\{\frac 1{(1\!+\!t)}\null_\frac\alpha2,\frac1{(1\!+\!|x|)}\null_\alpha\big\},\,t>0,\,|x|>\max\{2R_0,1\}.\ee
}\end{lemma}\Pr Properties \rf{HS0} are well known. To prove \rf{HSI}, we employ the representation formula and \rf{HSII}, so that
$$\ba{ll}|w(t,x)|\hskip-0.3cm&\leq\!\displ\intl{\R^3}\!\!H(t,x-y)|v_\circ(y)|dy=\!\!\!\!\!\intl{B(R_0)}\!\!\!\!\!H(t,x-y)|v_\circ(y)|dy\!+\!V_\circ\hskip-0.6cm\intl{\R^3\setminus B(R_0)}\hskip-0.6cmH(t,x-y)(1\!+\!|y|)^{-\alpha}dy\VSE\leq c(R_0)\dm v_\circ \dm_2(|x|+t^{\frac12})^{-3}
+cV_\circ\min\{(1\!+\!t)^{-\frac\alpha2},(1\!+\!|x|)^{-\alpha}\},\ea$$
provided that $|x|>\max\{2R_0,1\}$, which proves the lemma.\chiu
We also approach problem \rf{HS}   in
a form  weaker than the usual one for the initial
 value problem for the
Stokes equations. This weak
formulation, introduced in
\cite{MLI}, allows to consider
initial data in the Lebesgue spaces
$L^p,\,p\in[1,\infty]$, and not in the space of the
hydrodynamics $J^p, \,p\in(1,\infty)$.  Its interest is
connected with the possibility of
deducing  estimates in $L^r$-spaces with $r\in (1,\infty]$ by
means of duality arguments. Of
course, for an initial data in $J^p$
we come back to the classical
Stokes solutions.\par We have the
following special result, for a general formulation see \cite{MLI} (such a solution is denoted by $(\theta,\pi_\theta)$):
\begin{lemma}\label{LI}{\sl Let
$\theta_0\in C_0(\R^3).$ Then, to the data
$\theta_0$ there corresponds a unique smooth solution
$(\theta,c_0)$ to the Cauchy problem \rf{HS} such
that $\theta\in {\underset{q>1}\cap}C(0,T;
J^q(\R^3))$, $\theta\in {\underset{q>1}\cap}
L^q(\eta,T;W^{2,q}(\R^3)) $ and $\theta_t\in
{\underset{q>1}\cap}L^q(\eta,T;L^q(\R^3)),\,\eta>0$.
Moreover, for $q\in(1,\infty]$,
\be\label{HERA}\ba{lll}\dm \theta(t)\dm_q&\leq
c\dm \theta_0\dm_1t^{-\mu},& \mu\,=\frac 32
\big(1-\frac1q\big),\;t>0;\vspace{2pt}\\\dm\nabla
\theta(t)\dm_q\hskip-1cm&\leq c\dm \theta_0\dm_1t^{-\mu_1},&
\mu_1\hskip-0.1cm=\frac12+\mu\vspace{2pt}
\\\dm \theta_t(t)\dm_q&\leq c\dm \theta_0\dm_1t^{-\mu_2},\;
&\mu_2\hskip-0.1cm=1+\mu,\;t>0;\ea\ee with
$c$ independent of $\theta_0$. Finally,
${\underset{t\to0}\lim}(\theta(t)-\theta_0,\varphi)=0$
holds for any $\varphi\in C^1(\R^3)\cap
J^{1,q'}(\R^3)$.}\end{lemma}
\Pr See \cite{MLI}, Lemma\,3.2\,.\chiu
\begin{coro}\label{CLIcor}{\sl In the hypotheses of Lemma\,\ref{LI}, for all $\lambda\in(0,1]$, the following estimates hold:
\be\label{CLII}\ba{ll} &\dm \theta(t)-\theta(s)\dm_q\leq c\xi^{-\lambda-\frac32\left(1-\frac1q\right)}|t-s|^\lambda\dm \theta_0\dm_1,\VSE \dm \nabla \theta(t)-\nabla \theta(s)\dm_q\leq c\xi^{-\frac 32\lambda-\frac32\left(1-\frac1q\right)}|t-s|^\lambda\dm \theta_0\dm_1,\ea\ee
$t,s>0$, where $\xi:=\min\{s,t\}$.}\end{coro}\Pr Let $\xi:=\min\{s,t\}$. From \eqref{HERA}$_1$, $\|\theta(\frac{\xi}{2})\|_q\leq c\|\theta_0\|_1\xi^{-\mu}$. On the other hand, from the representation formula, one has 
$\theta(t,x)=\mathbb H[\theta(\frac{\xi}{2})](t-\frac\xi 2,x)$. Hence, using the $L^q$-H\o lder's properties of $\theta$, for all $\lambda\in (0,1]$,  we get
$$\dm \theta(t)-\theta(s)\dm_q\leq c\,\xi^{-\lambda}|t-s|^\lambda\dm \theta(\frac{\xi}{2})\dm_q\leq 
c\,\xi^{-\lambda-\frac32\left(1-\frac1q\right)}|t-s|^\lambda\dm \theta_0\dm_1\,.$$
Similar arguments lead to estimate \eqref{CLII}$_2$.\chiu
\section{\large A space time behavior of the Leray's generalized energy inequality} 
We start by proving  the following  interpolation inequality,
of the same kind of the one by
Gagliardo and Nirenberg.  
It is a particular case of a more general result for exterior domains, obtained in \cite{CMI}. 
The difference
with respect to the usual result is
 that the function $u$ does
not belong to a completion space of
$C^\infty_0(\R^3 )$. 
\begin{lemma}\label{ICM}{\sl  Let $u\in
W^{1,2}(\R^3\setminus B_R)$. Then there exists a
constant $c$ independent of $u$ and $R$ 
such that, for any $p\in [2,6]$,
\be\label{CMI} \dm u\dm_{L^p(|x|\geq R)}\leq \,c\dm \nabla 
u\dm_{L^2(|x|\geq R)}^a\dm u\dm_{L^2(|x|\geq R)}^{1-a}\,, \ a=\frac{3(p-2)}{2p}\,.\ee
}\end{lemma} \par
 \Pr Let $x\in \R^3\setminus B_R$ be the vertex of an infinite cone $C_x\subset \R^3\setminus B_R$, of fixed aperture independent of $x$. Let $(r,\theta)$ be spherical polar coordinates with origin at $x$, assume that the cone $C_x$ is given by $r\in (0,\infty)$ and $\theta\in \Theta$, and let $r^2\omega(\theta) drd\theta$ the volume element. Let $\{h_\rho(r)\}$ be a sequence of smooth cut-off functions such that $h_\rho(r)\in [0,1]$,  $h_\rho(r)=1$ for $r\leq \rho$, $h_\rho(r)=0$ for $r\geq 2\rho$, and $|h_\rho'(r)|\leq c\rho^{-1}$. Then 
 $$|u(x)|= |u(0,\theta)|=|-\int_0^\infty\!\!\frac{\partial}{\partial r}(u(r,\theta)h(r))dr|\leq\! \int_0^\infty\!\!|\nabla u(r,\theta)h(r)|dr+
\frac{c}{\rho}\!\int_0^\infty\!\!\!|u(r,\theta)|  dr.$$
Multiplying by $\omega(\theta) $ and integrating over $\Theta$ we get
$$4\pi|u(x)|\leq 
\int_{C_x}\frac{|\nabla u(y)|}{|x-y|^2}dy+\frac{c}{\rho}
\int_{C_x}\frac{|u(y)|}{|x-y|^2}dy
\leq \int_{|y|\geq R}\!\frac{|\nabla u(y)|}{|x-y|^2}dy+\frac{c}{\rho}\!
\int_{|y|\geq R}\!\frac{|u(y)|}{|x-y|^2}dy\,.$$
We let $\rho$ tend to infinity and then apply the Hardy-Littlewood-Sobolev theorem, and we find
\be\label{HLS}
\|u\|_{L^6(|x|\geq R)}\leq c\|\nabla u\|_{L^2(|x|\geq R)}\,,\ee
with $c$ independent of $R$.  
Using the interpolation between Lebesgue spaces 
$$\|u\|_{L^p(|x|\geq R)}\leq  c\|u\|_{L^6(|x|\geq R)}^a \|u\|_{L^2(|x|\geq R)}^{1-a},\  \ a=\frac{3(p-2)}{2p},$$
and then estimate \eqref{HLS}, we arrive at \eqref{CMI}. \chiu
 We set \be\label{AS1}u:=v-w\ \mbox{  and }\ \pi_u:=\pi_v, \ee where $(v,\pi_v)$ is a suitable weak solution to the Navier-Stokes Cauchy problem  and $w$ is the solution to the Stokes Cauchy problem. Both the solutions assume the initial data $v_\circ$. 
We define the symbol $\dm\cdot\dm_{L^p(k,R)}:=\dm\cdot\dm_{L^p(|y|>\frac k{k+1}R)}$, for all $k\geq0$.
\begin{lemma}\label{PO}{\sl In the hypotheses of Theorem\,\ref{CR}, for all $k\geq2$ and for $R>4R_0$, there exists a $c(k)$ such that \be\label{POI}\dm \pi_v\dm _{L^2(k,R)}\!\leq c\dm u\dm_{L^2(k-1,R)}^\frac12\dm \nabla u\dm_{L^2(k-1,R)}^\frac32 \!+\frac{V_\circ}{R^\alpha}\dm u\dm_{L^2(k-1,R)}\!+c{\frac{V_\circ^2}{R}\null_{\!2\alpha-\frac32}} \!+\frac1R\null_{\!\frac32}\dm v_\circ\dm_2^2,
\ee almost everywhere in $t>0$. }\end{lemma}
\Pr From \eqref{AS1} and the representation formula \rf{PWSI} of the pressure field  we get
\be\label{POII}\pi_v=\mathbb E(v\otimes v)=\mathbb E(u\otimes u)+\mathbb E(u\otimes w)+ \mathbb E(w\otimes u)+\mathbb E(w\otimes w)=\mbox{${\underset {i=1}{\overset4\sum}}$}\pi^i\,.\ee
For $i=1,\cdots,4$ and $|x|>\frac k{k+1}R$, we get, with obvious meaning of the symbols, the estimate
$$|\pi^i(x)|\leq c \intl{|y|<\frac{k-1}kR}\frac {|a||b|}{|x-y|}\null_3dy+\big|D_{x_hx_k}\intl{|y|>\frac{k-1}kR}\mathscr E(x-y)a_hb_k
dy\big|=:\pi^i_1+\pi^i_2\,.$$
By the assumption on $x$ there holds
$$|x-y|\geq |x|-|y|\geq \frac{|x|}{k^2},\mbox{ for all }|y|<\mbox{\large$\frac{k-1}k$}R,$$ and we get
$$|\pi^i_1(x)|\leq \frac{c(k)}{|x|^3}\intl{|y|<\frac{k-1}kR}|u||w|dy\leq \frac{c(k)}{|x|^3}\dm u\dm_{2}\dm w\dm_2,\, i=2,3\,, $$
which implies $$\dm\pi^i_1\dm_{L^2(k,R)}\leq \frac{c(k)}{R^\frac32}\dm u\dm_{2}\dm v_\circ\dm_2,\,i=2,3.$$
For the term $\pi^i_2$, by applying the Calder\'on-Zigmund theorem and then Lemma\,\ref{HS}, we get
$$\dm\pi^i_2\dm_{L^2(k,R)}\leq c\dm |u||w|\dm_{L^2(k-1,R)}\leq c\frac{V_\circ}{R^\alpha}\dm u\dm_{L^2(k-1,R)},\,i=2,3\,.$$
Repeating  the above arguments for the term $\pi^4$, we get
$$|\pi^4_1(x)|\leq \frac{c(k)}{|x|^3}\dm v_\circ\dm_2^2,$$ hence $$\dm \pi^4_1\dm_{L^2(k,R)}\leq \frac{c(k)}{R^\frac32}\dm v_\circ\dm_2^2.$$
Since $\alpha\geq1$ and $R>4R_0$, from Lemma\,\ref{LI} we easily deduce
$$\dm \pi^4_2\dm_{L^2(k,R)}\leq c\dm w^2\dm_{L^2(k-1,R)}\leq cV_\circ^2 R^{-2\alpha+\frac32}.$$
Finally, we estimate $\pi^1$. For $\pi^1_1$ we obtain the same estimate, that is
$$| \pi^1_1(x)|\leq \frac{c(k)}{|x|^3}\dm u\dm_2^2 \leq \frac{c(k)}{|x|^3}\dm v_\circ\dm_2^2.$$ Hence, we get
$$\dm \pi^1_1\dm_{L^2(k,R)}\leq \frac{c(k)}{R^\frac32}\dm v_\circ\dm_2^2.$$
Applying the Calder\'on-Zigmund theorem for singular integrals and estimate \rf{CMI}, we get
$$\dm \pi^1_2\dm_{L^2(k,R)}\leq c\dm u\dm_{L^4(k-1,R)}^2\leq c\dm u\dm_{L^2(k-1,R)}^\frac12\dm\nabla u\dm_{L^2(k-1,R)}^\frac32.$$ The above estimate and formula \rf{POII} give \rf{POI}.\chiu

\begin{lemma}\label{PO1}{\sl In the hypotheses of Theorem\,\ref{CR}, for all $k\geq2$ and for $R>2M_0R_0$, there exists a constant $c$ such that 
\be\label{CPE}\dm\pi_v\dm_{L^2(k,R)}\leq c\big(\dm v\dm_2^2+\dm v\dm_{L^4(k-1,R)}^2  \big),\ee almost everywhere in $t>0$. }\end{lemma}
\Pr
 Estimate \rf{CPE} is obtained by the same arguments used in the proof of Lemma\,\ref{PO}, provided that we consider $\mathbb E[v,v]$ and not its decomposition by means of $u,w$. The constant $c$ depends on $k$ and $R$, however it is bounded with respect either to $k$ and $R$.\chiu
\begin{lemma}\label{dens2} {\sl 
Assume that $(v,\pi_v)$ is a suitable weak solution. Then it satisfies the following inequality
 \be\label{SEIM}\ba{l}\displ\intl{\R^3}|v(t)|^2\psi(t)dx+2\intll st\intl
{\R^3}|\nabla v(\tau)|^2\psi \,dxd\tau\leq \intl{\R^3}|v(s)|^2\psi(s)dx\VS\hskip 2,5cm+
\intll st\intl{\R^3}|v|^2(\psi_\tau+\Delta\psi)dxd\tau+\intll st\intl{\R^3}(|v|^2+2\pi_v)
v\cdot\nabla\psi dxd\tau,\ea\ee 
for all $t\geq s$, for $s=0$ and a.e. in $s\geq 0$, and for all 
nonnegative $\psi\in C_0^\infty(\R; C_b^2(\R^3))$. 
}
\end{lemma}
\Pr 
 Taking into account the integrability properties of a suitable weak solution, all the terms in \eqref{SEIM} make sense. Let us define a sequence of smooth cut-off functions $\{h_\rho(x)\}$ with $h_\rho(x)\in  [0,1]$, $h_\rho(x)=1$ for $|x|\leq \rho$, $h_\rho(x)=0$ for $|x|\geq 2\rho$. Then for any $\eta>0$ and $\rho>0$ the function $\phi^{\rho,\eta}(t,x):=J_{\eta}(h_\rho(x)\psi(t,x))$, where $J_\eta$ is a Friedrichs mollifier,  belongs to $C_0^\infty(\R\times \R^3)$. Hence, from \eqref{SEI} written with $\phi$ replaced by $\phi^{\rho,\eta}$, passing to the limit as $\eta\to 0$ and, subsequently, as $\rho\to\infty$, by the integrability properties of $(v,\pi_v)$ and the Lebesgue dominated convergence theorem, we obtain the result.  
\chiu

\begin{lemma}\label{EIGF}{In the hypotheses of Theorem\,\ref{CR}, for all $t>0$ and for all nonnegative function $\varphi\in C_b^2(\R^3)$, such that $\varphi=0$ for $|x|\leq \max \{2R_0,1\}$, the following inequality holds 
\be\label{EIGFI}\ba{l}\displ \frac 12\intl{\R^3}|u(t)|^2\varphi dy+\intll 0t\intl{\R^3}|\nabla u|^2\varphi dyd\tau\leq \frac12
\intll0t\intl{\R^3}|u|^2\Delta\varphi dyd\tau+\intll0t\intl{\R^3}u\!\cdot\!\nabla \varphi\pi_udyd\tau\VS\hskip2cm+\frac12
\intll0t\intl{\R^3}\Big[|u|^2u\!\cdot\!\nabla \varphi +   |u|^2w\!\cdot\!\nabla\varphi\Big]dyd\tau\VS\hskip2.5cm+\intll0t\intl{\R^3}\Big[u\!\cdot\!\nabla u\!\cdot\! w\varphi+u\!\cdot\!\nabla\varphi u\!\cdot\! w+w\!\cdot\!\nabla u\!\cdot\! w\varphi+w\!\cdot\!\nabla\varphi u\!\cdot\! w\Big]dyd\tau .\ea\ee}\end{lemma}
\Pr The proof is an easy consequence of the Leray-Serrin technique. Indeed, from Lemma\,\ref{dens2}, we can consider the generalized energy inequality \eqref{SEIM} for a weak solution $v$  with $\psi(\tau,x):=h(\tau)\vp(x)$, with $\varphi\in C_b^2(\R^3)$ nonnegative, such that  $\varphi=0$ for $|x|\leq \max\{2R_0,1\}$, and $h\in [0,1]$ smooth cut-off function such that $h(\tau)=1$ for $\tau\in [s,t]$, $t>s>2\vep$, $h(\tau)=0$ for $\tau<\vep$ and in a neighborhood of $T$. Then the following inequality holds:
\be\label{SEI0} \ba{l}\displ\intl{\R^3}|v(t)|^2\varphi\,dx+2\intll st\intl
{\R^3}|\nabla v|^2\varphi dxd\tau\leq \intl{\R^3}|v(s)|^2\varphi dx\VS\hskip 2,5cm+\intll st\intl{\R^3}|v|^2\Delta\varphi dxd\tau+\intll st\intl{\R^3}(|v|^2+2\pi_v)
v\cdot\nabla\varphi  dxd\tau.\ea\ee Reasoning in analogous way for $w$, but recalling the regularity of $w$ and the linear character of the equations, we deduce
\be\label{SEII}\ba{l}\displ\intl{\R^3}|w(t)|^2\varphi dx+2\intll st\intl
{\R^3}|\nabla w|^2\varphi dxd\tau= \intl{\R^3}|w(s)|^2\varphi dx +\intll st\intl{\R^3}|w|^2\Delta\varphi dxd\tau .\ea\ee 
In the weak formulation of $(v,\pi_v)$ we can replace the test function $\varphi(\tau,x)$ by the function $\psi(\tau,x)h_\rho(x)w(\tau,x)=h(\tau)\varphi(x)h_\rho(x)w(\tau,x)$,
 with $\{h_\rho(x)\}\subset  [0,1]$ sequence of smooth cut-off functions, $h_\rho(x)=1$ for $|x|\leq \rho$, $h_\rho(x)=0$ for $|x|\geq 2\rho$. Then for any $\rho>0$ the function  $\psi(\tau,x)h_\rho(x)w(\tau,x)$ belongs to $C_0^1([0,T)\times \R^3)$. 
 Hence, we get 
$$\ba{l}\displ (v(t),
w(t)\varphi\, h_\rho)+\intll st\intl{\R^3}\nabla v\cdot\nabla w\varphi h_\rho dyd\tau=(v(s),w(s)\varphi h_\rho) \VS+\intll
st\Big[(v, w_\tau\varphi h_\rho)-(\nabla
v,\nabla
(\varphi h_\rho)\otimes w)+(v\cdot\nabla(\varphi h_\rho),v\cdot w)\VS \displ\hfill +(v\cdot\nabla w,\varphi h_\rho v)+(\pi_v,w\cdot\nabla
(\varphi h_\rho))\Big]d\tau.\ea$$
Recalling that $w_\tau=\Delta w$, and observing that, by interpolation, estimate \eqref{CCTI} and the energy inequality imply $v\in L^4(0,T; L^4(\R^3\setminus B_{M_0R_0}))$, we have
$$\ba{l}\displ (v(t),
w(t)\varphi h_\rho)+2\intll 
st\intl{\R^3}\nabla v\cdot\nabla w\varphi h_\rho dyd\tau=(v(s),w(s)\varphi h_\rho)\VS\hskip 1cm+\intll 
st\Big[(v,w\Delta(\varphi h_\rho))+(v\cdot\nabla(\varphi h_\rho),v\cdot w)+(v\cdot\nabla w,\varphi h_\rho v)+(\pi_v,w\cdot\nabla
(\varphi h_\rho))\Big]d\tau.\ea$$
By using the integrability properties of $v$ and $w$, passing to the limit as $\rho$ tends to infinity we find
$$\ba{l}\displ (v(t),
w(t)\varphi)+2\intll 
st\intl{\R^3}\nabla v\cdot\nabla w\varphi dyd\tau=(v(s),w(s)\varphi)
\VS\hskip 1.5cm+\intll 
st\Big[(v,w\Delta\varphi)+(v\cdot\nabla\varphi ,v\cdot w)+
(v\cdot\nabla w,\varphi v)+(\pi_v,w\cdot\nabla
\varphi)\Big]d\tau.\ea$$

Multiplying this last relation by $-2$ and summing to  the inequalities \rf{SEI0}-\rf{SEII},  recalling also that  $\varphi $ is null for $|x|\leq\max\{2R_0,1\}$ ensures that $w(t,x)$ satisfies estimate \eqref{HSI}, we deduce
\be\label{temp}\ba{l}\displ \frac 12\intl{\R^3}|u(t)|^2\varphi dy+\intll st\intl{\R^3}|\nabla u|^2\varphi dyd\tau\leq \frac12
\intl{\R^3}|u(s)|^2\varphi dy+ \frac12
\intll st\intl{\R^3}|u|^2\Delta\varphi dyd\tau\VS\hskip2cm+\intll st\intl{\R^3}u\!\cdot\!\nabla \varphi\pi_udyd\tau+\frac12
\intll st\intl{\R^3}\Big[|u|^2u\!\cdot\!\nabla \varphi +   |u|^2w\!\cdot\!\nabla\varphi\Big]dyd\tau\VS\hskip2cm+\intll st\intl{\R^3}\Big[u\!\cdot\!\nabla u\!\cdot\! w\varphi+u\!\cdot\!\nabla\varphi u\!\cdot\! w+w\!\cdot\!\nabla u\!\cdot\! w\varphi+w\!\cdot\!\nabla\varphi u\!\cdot\! w\Big]dyd\tau .\ea\ee
Thanks to the integrability properties of $u$ and the regularity of $w$ for $|x|>2R_0$, given in Lemma\,\ref{HS}, we can pass to the limit as $s>2\vep $ tends to $0$ and we get \rf{EIGFI}.\chiu
This lemma is a relevant tool for proving Lemma\,\ref{ELI} on the asymptotic behavior in $R$ of the $L^2$-norm of $u=v-w$ outside the ball $B_R$, uniformly in $\alpha$. Actually, if we start from the energy inequality, we could get an asymptotic behavior in $R$ only for $\alpha>\frac 32$.

 Assume that $\varphi\in C_b^2(\R^3)$ is defined as follows 
  \be\label{CO} \ba{ll}
 m\geq2, &\displ \;\varphi\!:=\varphi_R(m)\!:=\!\left\{\!\!\ba{ll}1& \!\!\mbox{if }|x|\geq \frac m{m+1} R,\\\in [0,1]&\!\!\mbox{if }|x|\in[\frac{m-1}mR,\frac m{m+1}R],\\ 0&\!\!\mbox{if }|x|\leq \frac{m-1}mR,\ea\right.\; \VS
 &\displ \!|\nabla\varphi_R(m)|+R|\Delta \vp_R(m)|\leq c(m)R^{-1}.\ea\ee
We define $\phi_R(m)(\tau,x):=k(\tau)\,\vp_R(m)(x)$ 
with $k(\tau)\in [0,1]$ smooth cut-off function such that $k(\tau)=1$ for $|\tau|\leq t$ and $
k(\tau)=0$ for $|\tau|\geq 2t$.  
Since $\phi_R(m)$ belongs to $C_0^\infty(\R; C_b^2(\R^3))$,  by Lemma\,\ref{dens2} we can use $\phi_R(m)$ as test function in \rf{SEI}, and we get the following generalized energy inequality \eqref{SEIi}, that we will call generalized Leray's energy inequality:
\be\label{SEIi}\ba{l}\displ\intl{\R^3}|v(t)|^2\vp_R(m)dx+2\intll st\intl
{\R^3}|\nabla v|^2\vp_R(m)dxd\tau \leq \intl{\R^3}|v(s)|^2\vp_R(m)dx\VS\hskip 2,5cm+\intll st\intl{\R^3}|v|^2\Delta\vp_R(m)dxd\tau+\intll st\intl{\R^3}(|v|^2+2\pi_v)
v\cdot\nabla\vp_R(m) dxd\tau,\ea\ee for all $t\geq s$, for $s=0$ and a.e. in $s\geq 0$. 

\begin{lemma}\label{ELI}{\sl The following estimate holds
\be\label{ELII}\dm u(t)\dm_{L^2(6,R)}\leq cR^{-\frac12}t^\frac12C(v_\circ),\;\, R>4R_0, t>0,\ee
with $C^2(v_\circ):=V_\circ^4+V_\circ^2\dm v_\circ\dm_2^2+V_\circ^2\dm v_\circ\dm_2+V_\circ\dm v_\circ\dm_2^2+\dm v_\circ\dm_2^2+\dm v_\circ\dm_2^3$.} \end{lemma}
\Pr We consider the sequence of function \rf{CO}. In \rf{EIGFI} we replace   $\varphi$ by $\varphi_R^2(k)$. By obvious meaning of the symbols on the right-hand side, we obtain
$$\frac 12\dm u(t)\varphi_R(k)\dm_2^2+\intll 0t\intl{\R^3}|\nabla u|^2\varphi_R^2(k) dyd\tau\leq\mbox{${\overset 8{\underset{i=1} \sum}}$}|I_i(t,k)|.$$
We estimate each $I_i(t,k),i=1,\dots,8$. Recalling the definition of $\varphi_R(k)$, we get
\begin{itemize}\item
$\displ|I_1|\leq c(k)R^{-2}\intll0t\dm u\dm_{L^2(k-1,R)}^2d\tau;$\item by virtue of estimate \rf{POI} applying H\"older's inequality, we get $$\ba{ll}\displ
|I_2|\hskip-0.2cm&\displ\leq c(k)R^{-1}\!\!\intll0t\! \dm |\pi_u||u|\dm_{L^1(k-1,R)}d\tau\leq 
c(k)R^{-1}\!\!\intll0t \!\dm \pi_u\dm_{L^2(k-1,R)}\dm u\dm_{L^2(k-1,R)}d\tau \VSE
\leq c(k)R^{-1}\!\!\intll0t\Big[\!\dm u\dm_{L^2(k-2,R)}^\frac12\dm \nabla u\dm_{L^2(k-2,R)}^\frac32 d\tau+\frac{V_\circ}{R^\alpha}\dm u\dm_{L^2(k-2,R)}\VSE\hskip3cm+c{\frac{V_\circ^2}{R}\null_{\!2\alpha-\frac32}} +\frac1R\null_{\!\frac32}\dm v_\circ\dm_2^2\Big]\dm u\dm_{L^2(k-1,R)}d\tau\,.
\ea$$
\item by virtue of estimate \rf{CMI}, we get
$$|I_3|\leq c(k)R^{-1}\intll0t\dm u\dm_{L^3(k-1,R)}^3d\tau\leq c(k)R^{-1}\intll0t\dm u\dm_{L^2(k-1,R)}^\frac32\dm\nabla u\dm_{L^2(k-1,R)}^\frac32d\tau;$$
\item by virtue of \rf{HSI}, we have  $$ |I_4+I_6|\leq c(k)R^{-1-\alpha}V_\circ\intll0t\dm u\dm_{L^2(k-1,R)}^2d\tau ;$$
\item by virtue of \rf{HSI}, applying first the H\"older inequality and then the Cauchy inequality, we get
$$\ba{ll}|I_5|\hskip-0.2cm&\displ\leq cR^{-2\alpha}V_\circ^2\intll0t \dm u\varphi_R(k)\dm_2^2d\tau+\frac14\intll0t\dm\varphi_R(k)\nabla u\dm_2^2d\tau\VSE \leq cR^{-2\alpha}V_\circ^2\intll0t \dm u\dm_{L^2(k-1,R)}^2d\tau+\frac14\intll0t\dm\varphi_R(k)\nabla u\dm_2^2d\tau;\ea$$
\item by virtue of \rf{HSI}, applying the H\"older inequality and then the Cauchy inequality, we get
$$\ba{ll}|I_7|\hskip-0.2cm&\displ\leq c\intll0t \dm w\varphi_R(k)\dm_4^4d\tau+\frac14\intll0t\dm\varphi_R(k)\nabla u\dm_2^2d\tau\VSE \leq cV_\circ^4R^{-4\alpha+3}t+\frac14\intll0t\dm\varphi_R(k)\nabla u\dm_2^2d\tau;\ea$$
\item by virtue of \rf{HSI}, applying the H\"older inequality, we get
$$|I_8|\leq c(k)R^{-1}\intll0t\dm w\dm_{L^4(k-1,R)}^2\dm u\dm_{L^2(k-1,R)}d\tau\leq c(k)R^{-2\alpha+\frac12}V_\circ^2\intll0t\dm u\dm_{L^2(k-1,R)}d\tau.
$$  \end{itemize}
The above estimates allow to deduce the following one:
\be\label{ELIIV}\ba{l}\displ\frac 12\dm u(t)\varphi_R(k)\dm_2^2+\frac12\intll 0t\intl{\R^3}|\nabla u|^2\varphi_R^2(k) dyd\tau\VS\leq c(k)R^{-1}\Big[ C_0(v_\circ)t \;+
\intll0t\dm u\dm_{L^2(k-2,R)}^\frac32\dm \nabla u\dm_{L^2(k-2,R)}^\frac32d\tau\Big],\ea
\ee
with $C_0(v_\circ):= V_\circ^4+V_\circ^2\dm v_\circ\dm_2^2+V_\circ^2\dm v_\circ\dm_2+V_\circ\dm v_\circ\dm_2^2+\dm v_\circ\dm_2^2$. 
Writing estimate \rf{ELIIV} with $k=2$ gives 
\be\label{ELIVI}\displ\frac 12\dm u(t)\varphi_R(2)\dm_2^2\!+\!\frac12\!\intll 0t\!\intl{\R^3}\!\!|\nabla u|^2\varphi_R^2(2) dyd\tau\!\leq\!c(2)R^{-1}(C_0(v_\circ)\!+\!\dm v_\circ\dm_2^3)t^\frac14,\mbox{ for }t\in(0,1),\ee and 
\be\label{ELIVIX}\displ\frac 12\dm u(t)\varphi_R(2)\dm_2^2+\frac12\!\intll 0t\intl{\R^3}|\nabla u|^2\varphi_R^2(2) dyd\tau\leq c(2)R^{-1}(C_0(v_\circ)+\dm v_\circ\dm_2^3)t,\mbox{ for }t>1,\ee 
which proves \rf{ELII} for $t\geq1$. Taking into account estimate \rf{ELIVI}, we evaluate \rf{ELIIV} for $k=4$ and $t\in(0,1)$, and  we get
\be\label{ELIVII}\displ\frac 12\dm u(t)\varphi_R(4)\dm_2^2+\frac12\intll 0t\intl{\R^3}|\nabla u|^2\varphi_R^2(4) dyd\tau\leq c(4)R^{-1}t^\frac58(C_0(v_\circ)+\dm v_\circ \dm_2^3),\mbox{ for }t\in(0,1).\ee 
Taking  into account \rf{ELIVII} and evaluating \rf{ELIIV} for $k=6$ and $t\in (0,1)$, we get 
$$\frac 12\dm u(t)\varphi_R(6 )\dm_2^2+\frac12\intll 0t\intl{\R^3}|\nabla u|^2\varphi_R^2(6) dyd\tau\leq c(6)R^{-1}(C_0(v_\circ)t+\dm v_\circ \dm_2^3t^\frac{19}{16}),\mbox{ for }t\in(0,1),$$ which gives
\rf{ELII} for $t\in(0,1)$. This last estimate and \rf{ELIVIX} complete the proof.\chiu
\section{\large{Pointwise representation of the weak solution for {\normalsize$(t,x)\in(0,T)\times(\R^3\setminus B_{2R_0})$}}}
The following result is similar to Lemma 3.5 in \cite{MDCDS}, but we are replacing a $J^2$-continuity assumption with a $J^2$-weak continuity one. \begin{lemma}\label{lDCDS}{\sl Let  $v(t)$ be a $J^2(\R^3)$-weakly continuous function on $[0,\!T)$, $\psi\!\in\!C(0,\!T;\!J^2(\R^3)\!)$, and $\psi_0 \in L^2(\R^3)$ such that $\lim_{t\to 0} (\psi(t),\vp)=(\psi_0,\vp)$ for all $\vp\in J^{1,2}(\R^3)$. Then, for all $t\in (0,T)$, the following limit property holds
 $$\lim_{\delta\to 0}(v(t-\delta),\psi(\delta))=(v(t),\psi_0)\,.$$  }  
 \end{lemma}
\Pr From the limit assumption on $\psi(t)$ as $t\to 0$ and from  the $J^2(\R^3)$-strong continuity of $\psi(t)$, we can infer that $\psi(0)=P(\psi_0)$. 
Since for any $t\in [0,T)$,  $v(t)\in J^2(\R^3)$, for any $\delta>0$ we have
$$\ba{ll}|(v(t-\delta), \psi(\delta))-(v(t),\psi_0)|&= |(v(t-\delta)-v(t), \psi_0)+(v(t-\delta), \psi(\delta)-\psi_0)|\VS &\displ\leq 
|(v(t-\delta)-v(t), \psi(0))|+\|v(t-\delta)\| \|\psi(\delta)-\psi(0)\|\,.\ea$$
Using the $J^2$-weak continuity for the first term on the right-hand side, and the  $J^2$-strong continuity of $\psi(\delta)$ for second one, we get the result.\chiu
\par
 We premise the following regularity result of the weak solution $v(t,x)$:
\begin{lemma}\label{CLI}{\sl In the hypotheses of Theorem\,\ref{CR} we get
$v(t,x)\!\in\!C(0,T;L^\infty(\R^3\setminus \ov {B}_{2M_0R_0}))$.}\end{lemma}
\Pr Let $\{h_\eta\}\subset [0,1]$ be a sequence of smooth cut-off functions with $h_\eta(t-\tau)=1$ for $t-\tau>2\eta$, $h_\eta(t-\tau)=0$ for $t-\tau<\eta$.  Let us consider the Navier-Stokes weak formulation corresponding to a solution $(v,\pi_v)$ written on the interval $(0,t-2\eta)$, with 
 $h_\eta(t-\tau)\varphi_R(4)\theta ^t(\tau,x)$ as test function.  The function $\varphi_R(4)$ is defined in \rf{CO},  and we set $R:=2 M_0R_0$, while $\theta^t:=\theta(t-\tau,x)$, for $\tau\in(0,t)$, with $\theta(\sigma,x)$ solution to the Stokes Cauchy problem \rf{HSO} given in Lemma\,\ref{LI}. It is known that $\theta^t$ is a backward in time solution to the Stokes Cauchy problem  on $(0,t)\times\R^3$.
 In the following, since there is no danger confusion, we denote $\varphi_R(4)$ simply by $\varphi$. \par Hence, after substituting, we get 
\be\label{eta1}\ba{l}\displ(v(t-2\eta),\varphi\theta(2\eta))= (v_\circ, \varphi \theta(t))+ 2\intll0{t-2\eta} (v,\nabla \varphi\cdot\nabla\theta^t)d\tau+\intll0{t-2\eta} (v,\theta^t\Delta \varphi)d\tau\VS\hskip 4.5cm- \intll0{t-2\eta} (v\cdot\nabla v, \varphi\theta^t)d\tau+\intll0{t-2\eta}(\pi_v,\nabla\varphi\cdot\theta^t)d\tau.\ea\ee
The same relation written on the interval $(0,s-2\eta)$ and with the test function $h_\eta(s-\tau)\varphi_R(4)\theta ^s(\tau,x)$ furnishes
\be\label{eta2}\ba{l}\displ(v(s-2\eta),\varphi\theta(2\eta))= (v_\circ, \varphi \theta(s))+2\intll0{s-2\eta} (v,\nabla \varphi\cdot\nabla\theta^s)d\tau+\intll0{s-2\eta}(v,\theta^s\Delta \varphi)d\tau\VS\hskip 4.5cm- \intll0{s-2\eta} (v\cdot\nabla v, \varphi\theta^s)d\tau+\intll0{s-2\eta}(\pi_v,\nabla\varphi\cdot\theta^s)d\tau.\ea\ee
From Lemma\,\ref{lDCDS}  one has
$$\lim_{\eta \to 0} (v(t-2\eta), \theta(2\eta))=(v(t),\theta_0), \mbox{ and }  \lim_{\eta \to 0} (v(s-2\eta), \theta(2\eta))=(v(s),\theta_0).$$
Passing to the limit as $\eta$ tends to $0$ on the right-hand side of \eqref{eta1} and \eqref{eta2},  we can use the Lebesgue dominated convergence theorem, observing that  the integrals on $(0,t)$, such as the integrals on $(0,s)$, are finite thanks to the estimates \eqref{HERA} with $q<\frac 32$. 
Hence, we get 
$$\ba{l}\displ(v(t),\varphi\theta_0)= (v_\circ, \varphi \theta(t))+2\intll0t (v,\nabla \varphi\cdot\nabla\theta^t)d\tau+\intll0t (v,\theta^t\Delta \varphi)d\tau\VS\hskip 4.5cm- \intll0t (v\cdot\nabla v, \varphi\theta^t)d\tau+\intll0t(\pi_v,\nabla\varphi\cdot\theta^t)d\tau,\ea$$
and 
$$\ba{l}\displ(v(s),\varphi\theta_0)= (v_\circ, \varphi \theta(s))+ 2\intll0s (v,\nabla \varphi\cdot\nabla\theta^s)d\tau+\intll0s(v,\theta^s\Delta \varphi)d\tau\VS\hskip 4.5cm- \intll0s (v\cdot\nabla v, \varphi\theta^s)d\tau+\intll0s(\pi_v,\nabla\varphi\cdot\theta^s)d\tau.\ea$$
Then, we deduce
\be\label{CLIrel}\ba{l}\displ(v(t)-v(s),\varphi\theta_0)= (v_\circ, \varphi(\theta(t)- \theta(s)))+2\intll st (v,\nabla \varphi\cdot\nabla\theta^t) d\tau+\intll{s}{t}(v,\theta^t\Delta \varphi)d\tau\VS\hskip 4.5cm- \intll st (v\cdot\nabla v, \varphi\theta^t)d\tau+\intll st(\pi_v,\nabla\varphi\cdot\theta^t)d\tau  \VS\hskip2.5cm +
\mbox{${\overset3{\underset{i=1}\sum}}$}\Big[2\intll{s_i}{s_{i+1}}(v,\nabla \varphi\cdot(\nabla\theta^t-\nabla\theta^s))d\tau+\intll{s_i}{s_{i+1}}(v,(\theta^t-\theta^s)\Delta \varphi)d\tau\VS\hskip 3cm-  \intll{s_i}{s_{i+1}}(v\cdot\nabla v, \varphi(\theta^t-\theta^s))d\tau+\intll{s_i}{s_{i+1}}(\pi_v,\nabla\varphi\cdot(\theta^t-\theta^s))d\tau \Big],\ea\ee
with $s_1=0,s_2=\vep,s_3=s-\vep$ and $s_4=s$.
We limit ourselves to estimate the first term, all the integrals involving the nonlinear term $v\cdot\nabla v$ and the pressure field, as the others are of simpler discussion. \begin{itemize}\item[i)] 
Applying H\"older's inequality and the semigroup properties \rf{CLII} of the solution $\theta$, we get
$$|(v_\circ, \varphi(\theta(t)- \theta(s)))| \leq \dm  v_\circ\dm_2\dm \theta(t)-\theta(s)\dm_2\leq c\dm v_\circ\dm_2\xi^{-1-\frac34}(t-s)\dm \theta_0\dm_1;$$
\item[ii)] applying H\"older's inequality and the semigroup properties (\ref{HERA}), recalling estimate \rf{CCTI} for the weak solution $v$, for $p=6$, we get
$$\ba{ll}\displ\Big| \intll st (v\cdot\nabla v, \varphi\theta^t)d\tau\Big|\hskip-0.2cm&
\displ\leq 
\intll st \big[\dm|v|^2|\nabla \varphi||\theta^t|\dm_1+\dm|v|^2\varphi|\nabla\theta^t|\dm_1\big]d\tau\VSE
\leq \!\intll st\!\!\big[ \dm v\dm^2_{L^{2p}(4,R)}\dm\theta^t\dm_{p'}\!+\dm v\dm^2_{L^{2p}(4,R)}\dm\nabla\theta^t\dm_{p'}\big] d\tau
\VSE
\leq \!\intll st\!\!\big[
\dm v\dm^{\frac2p}_{L^{2}(4,R)}  \dm v\dm^{\frac{2}{p'}}_{L^{\infty}(4,R)}(\dm\theta^t\dm_{p'}+\dm\nabla\theta^t\dm_{p'})\big] d\tau\VSE
\leq \dm v_\circ\dm_2^\frac13c(v_\circ)^\frac53s^{-\frac56}\big[(t-s)^{\frac34}+(t-s)^{\frac14}\big]\dm\theta_0\dm_1\,;
\ea$$
\item[iii)] applying H\"older's inequality, the semigroup properties \rf{HERA} of $\theta$ and taking into account that $t-\tau>s-\tau$, we obtain $$\ba{ll}\displ\big|\intll{s_1}{s_{2}}(v\cdot\nabla v, \varphi(\theta^t-\theta^s))d\tau\big|
\displ\leq  
\intll{s_1}{s_{2}}\dm v\cdot\nabla v \theta^t\dm_1+\dm v\cdot\nabla v\theta^s\dm_1d\tau 
\VS\hskip4cm
\leq \intll{s_1}{s_{2}}\dm v\dm_2\dm\nabla v \dm_2\dm\theta^t\dm_\infty+\dm v\dm_2\dm\nabla v\dm_2\dm\theta^s\dm_\infty d\tau 
\VS\hfill \leq 
c\dm v_\circ\dm_2\Big[\intll0\vep (s-\tau)^{-3}d\tau\Big]^\frac12\Big[\intll0\vep\dm\nabla v\dm_2^2d\tau\Big]^\frac12\dm \theta_0\dm_1
\leq 
c\dm v_\circ\dm_2^2\vep^\frac12\frac{(2s\!-\!\vep)^{\frac12}}{s(s-\vep)}\dm\theta_0\dm_1;
\ea$$
\item[iv)] applying H\"older's inequality and the semigroup properties \rf{CLII} of $\theta$, recalling estimate \rf{CCTI} for $v$ and that $s-\tau<t-\tau$, for $p=6$, we get
 $$\ba{ll}\displ\big|\intll{s_2}{s_{3}}(v\cdot\nabla \varphi, v\cdot(\theta^t-\hskip-0.2cm&\theta^s))d\tau\big| \displ+\big|\intll{s_2}{s_{3}}(v\cdot\nabla (\theta^t-\theta^s),v\varphi)d\tau\big|\VSE\leq  
\intll{s_2}{s_{3}}\Big[\dm v\dm_{2p}^2\dm\nabla ( \theta^t-\theta^s)\dm_{p'}+\dm v\dm_{2p}^2\dm\theta^t-\theta^s\dm_{p'}d\tau\Big] \VSE\leq c\dm v_\circ\dm_2^\frac2{p}c(v_\circ)^\frac2{p'}\dm\theta_0\dm_1
(t\!-\!s)\!\!\intll{\vep}{s-\vep} \frac{\ov\xi^{-\frac32-\frac3{2p}}\!+\xi^{-1-\frac 3{2p}}}{\tau^{\frac1{p'}}} d\tau\VSE  \leq 
c\dm v_\circ\dm_2^{\frac13}c(v_\circ)^\frac53\big[\vep^{-\frac{19}{12}}+\vep^{-\frac{13}{12}}\big](t-s)\dm\theta_0\dm_1;
\ea$$ 
\item[v)] applying H\"older's inequality and the semigroup properties \rf{HERA}, recalling estimate \rf{CCTI} for the weak solution $v$ and that $s-\tau<t-\tau$, for $p=6$, we get
$$\ba{ll}\displ\big|\intll{s_3}{s_{4}}(v\cdot\nabla \varphi,v\cdot(\theta^t-\hskip-0.2cm&\displ\theta^s))d\tau\big|+
\big|\intll{s_3}{s_{4}}(v\cdot\nabla (\theta^t-\theta^s),v\varphi)d\tau\big|
  \VSE\leq \intll{s_3}{s_{4}}\dm v\dm_{L^{2p}(4)}^2\dm\nabla (\theta^t-\theta^s)\dm_{p'} +\dm v\dm_{L^{2p}(4)}^2\dm\theta^t-\theta^s\dm_{p'} d\tau\VSE \leq 
c\dm v_\circ\dm_2^\frac2{p}c(v_\circ)^\frac2{p'}\Big[\intll{s-\vep}s \!\frac{\dm\nabla\theta^t\dm_{p'}\!+\dm\nabla\theta^s\dm_{p'}\!+\dm\theta^t\dm_{p'}\!+\dm\theta^s\dm_{p'}}{\tau^{\frac1{p'}}}d\tau\Big] \VSE \leq 
c\dm v_\circ\dm_2^\frac13c(v_\circ)^\frac53\frac{\vep^{\frac1{4}}+\vep^{ \frac3{4}}}{(s-\vep)^\frac 56}
 \dm\theta_0\dm_1.
\ea$$
\end{itemize}
Finally, we estimate the terms with the pressure field $\pi_v$.
To this end, taking into account that, by definition, $\nabla\varphi$ is nonnull on $B_{\frac 43 R}\setminus B_{\frac 34 R}$,  we can use estimate \rf{CPE}  for $k=3$:
$$\dm \pi_v\dm_{L^2(3,R)}\leq c\big(\dm v\dm_2^2+\dm v\dm_{L^4(2,R)}^2\big),\mbox{ a. e. in }t>0.$$
By using \rf{CCTI} and the energy relation, we get
\be\label{CEP}\dm \pi_v\dm_{L^2(3,R)}\leq c\big(\dm v_\circ\dm_2^2+\dm v_\circ\dm_2c(v_\circ)t^{-\frac12}\big).\ee
We start with the estimate on $(s,t)$:
\begin{itemize}\item[j)] applying H\"older's inequality, the semigroup properties \rf{HERA} for $\theta$ and estimate \rf{CEP}, we deduce
$$ \big|\intll st(\pi_v\nabla \varphi,\theta^t)d\tau\big|\leq \intll st \dm\pi_v\dm_{L^2(3,R)}\dm \theta^t\dm_2d\tau \leq c  \big(\dm v_\circ\dm_2^2+\dm v_\circ\dm_2c(v_\circ)s^{-\frac12}\big)(t-s)^\frac14\dm \theta_0\dm_1;$$
\item[jj)] by the same arguments as before and taking into account that $s-\tau<t-\tau$, we get
$$\ba{ll}\displ\big|\intll{s_1}{s_{2}}(\pi_v,\nabla\varphi\cdot(\theta^t-\theta^s))d\tau\big|\hskip-0.2cm&\displ\leq c \intll0\vep \dm\pi_v\dm_{L^2(3,R)}(\dm \theta^t\dm_2+\dm \theta^s\dm_2)d\tau\VSE\leq \frac c{(s-\vep)}\!\null_\frac34\dm\theta_0\dm_1\!\!\intll0\vep\big(\dm v_\circ\dm_2^2+\dm v_\circ\dm_2c(v_\circ)\tau^{-\frac12}\big)d\tau\VSE\leq  \frac c{(s-\vep)}\!\null_\frac34  \big(\dm v_\circ\dm_2^2\vep^\frac12+\dm v_\circ\dm_2c(v_\circ)\big)\vep^\frac12\dm \theta_0\dm_1;\ea$$
\item[jjj)] by the same arguments and employing estimate \rf{CLII}, we get
$$\ba{ll}\displ\big|\intll{s_2}{s_{3}}(\pi_v,\nabla\varphi\cdot(\theta^t-\theta^s))d\tau\big|\hskip-0.2cm&\displ\leq c \intll\vep{s-\vep} \dm\pi_v\dm_{L^2(3,R)}\dm \theta^t- \theta^s\dm_2d\tau\VSE\leq c\big(\dm v_\circ\dm_2^2+\dm v_\circ\dm_2c(v_\circ)\vep^{-\frac12}\big)\!\!\intll\vep{s-\vep}
\dm \theta^t-\theta^s\dm_2 d\tau\VSE\leq    c\big(\dm v_\circ\dm_2^2+\dm v_\circ\dm_2c(v_\circ)\vep^{-\frac12}\big)
\vep^{-\frac74}(t-s)\dm \theta_0\dm_1;\ea$$
\item[jv)] finally,
$$\ba{ll}\displ\big|\intll{s_3}{s_{4}}(\pi_v,\nabla\varphi\cdot(\theta^t-\theta^s))d\tau\big|\hskip-0.2cm&\displ\leq c \intll{s-\vep}s \dm\pi_v\dm_{L^2(3,R)}\dm \theta^t- \theta^s\dm_2d\tau\VSE\leq c\big(\dm v_\circ\dm_2^2+\dm v_\circ\dm_2c(v_\circ)(s-\vep)^{-\frac12}\big)\!\!\intll{s-\vep}s
\big(\dm \theta^t\dm_2+\dm\theta^s\dm_2\big) d\tau\VSE\leq    c\big(\dm v_\circ\dm_2^2+\dm v_\circ\dm_2c(v_\circ)(s-\vep)^{-\frac12}\big)
\vep^{\frac14}\dm \theta_0\dm_1.\ea$$
\end{itemize}
Considering estimates i)-v) and j)-jv), and the corresponding estimates for the linear terms of relation \rf{CLIrel}, we deduce $$|(v(t)-v(s),\varphi\theta_0)| \leq F(\vep,s,t,v_\circ)\dm \theta_0\dm_1,\mbox{ for arbitrary }\theta_0\in C_0(\R^3),$$
with clear meaning of function $F$, 
hence
$$\dm (v(t)-v(s))\varphi\dm_\infty\leq F(\vep,s,t,v_\circ).$$
Since we can assume $t-s<\vep^{4}$, for fixed $s>0$ and $v_\circ$ the above estimates ensure $\displ \lim_{\vep\to0}F(\vep,s,t,v_\circ)=0$, 
the lemma is proved.\chiu
In the following lemma we give the representation formula for the weak solution $(v,\pi_v)$ provided that $|x|>2M_0R_0$. To this end, we recall that the representation formula is given by means of the fundamental heat kernel $H$ (for its definition and properties see \rf{HS00}-\rf{HSII}) and the Oseen tensor $T$ (see \cite{K}):
$$\ba{rl} \vs1 T_{ij}(t-\tau,x-y):=&\dy-\Delta\phi(t-\tau,\vert
x-y\vert)\delta_{ij}+D^2_{x_ix_j}\phi(t-\tau,\vert
x-y\vert)\\
=&\dy H_{ij}(t-\tau,
x-y)+D^2_{x_ix_j}\phi(t-\tau,\vert
x-y\vert),\ea$$
$$\phi(t,r)=\frac
12\frac1{\sqrt\pi}\null_{\! \null_3}\frac
1r\intl 0^{r/2\sqrt t}e^{-\rho^2}d\rho.$$
We denote by 
$T_j(t-\tau,x-y)$ the j-{th} column of the matrix $T_{ij}$. The pair
$(T_j(t-\tau,x-y),p)$, with $p=0$, for 
$t-\tau>0$ is a solution in the $(t,x)$ variables of the Stokes system,
 and in the $(\tau,y)$ variables of its adjoint system:
$$\widehat w_\tau+\Delta\widehat w+\nabla p=0,\quad
\nabla\cdot \widehat w=0.$$ 
The following estimate holds:
\be\label{TPE} |D^k_sD^\beta_zT(s,z)|
\leq c(|z|+s^\frac12)^{-3-2k-|\beta|}\,.\ee Finally, we recall that from the definition of $T$ we get
$$(T_j(t,x),\varphi)=(H_j(t,x),\varphi), \mbox{ for all }\varphi\in J^p(\R^3),\;p>1.$$
We set $$ \mathbb T[v\otimes v](s,t,x):=\intll st (v\cdot \nabla T_i(t,x),v)d\tau\,.$$
If $s=0$, we simply write $\mathbb T[v\otimes v](t,x).$ \begin{lemma}\label{RF}{\sl Let $(v,\pi_v)$ be a suitable weak solution to the Navier-Stokes equations. Then, for all $t>0$ and $s\geq 0$, and almost a.e. in $x\in \R^3\setminus  B_{2M_0R_0}$, 
\be\label{RFI} v(t,x)=\mathbb H[v(s)](t-s,x)+ \mathbb T[v\otimes v](s,t,x)\,.\ee
Moreover, for $t>T_0$ and for $x\in \R^3$,
\be\label{RFIa} v(t,x)=\mathbb H[v(T_0)](t-T_0,x)+ \mathbb T[v\otimes v](T_0,t,x)\,.\ee
}\end{lemma}
\Pr Let us consider the weak formulation with  a divergence free test function:
\be\label{LII}
\ba{l}\displ(v(t),\varphi(t))= (v(s), \varphi(s))- \intll st 
(v,\varphi_\tau+\Delta \varphi)d\tau+ \intll st (v\cdot\nabla \varphi,v)d\tau.\ea\ee
We set, for all $i=1,2,3$, and $t-\vep>s$, 
$$\varphi(\tau,y):=h^\vep(\tau)J_n[T_i(t-\tau,x)](y),\mbox{ with }h^\vep(\tau):=\left\{\ba{ll}1&\mbox{if }\tau\leq t-\vep,\\ \in [0,1]&\mbox{if }\tau\in[t-\vep,t-\frac\vep4],\\ 0&\mbox{if }\tau>t-\frac\vep4,\ea\right.$$
where $J_n$ is 
a Friedrichs mollifier. Hence, inserting such a $\varphi$ in \eqref{LII} with $t$ replaced by $t-\vep$, and recalling that $T_i$ is a solution backward  in time with respect to $(\tau,y)\in (0,t)\times\R^3$, we obtain
\be\label{PRF}(v_i(t-\vep), J_n[H(\vep,x)])=(v_i(s),J_n[H(t-s,x)])
+\intll s{t-\vep}(v\cdot \nabla J_n[T_i(t-\tau,x)],v)d\tau.
\ee
We perform the limit as $\vep\to0$. To this end, we deal separately with  the terms of the last integral equation. 
\par i) The first term can be written as:$$\ba{ll}(v_i(t\!-\!\vep), J_n[H(\vep,x)])\!-\! J_n[v_i(t)](x)\VS \displaystyle\hskip3cm =\intl{\R^3}\!\!v_i(t\!-\!\vep,y)\!\!\intl{\R^3}\!\!J_n(y\!-\!z)H(\vep,x\!-\!z)dzdy-\!\!\intl{\R^3}\!\!J_n(x\!-\!y)v_i(t,y)dy\VS\hfill=\intl{\R^3}H(\vep,x-z)\Big[\intl{\R^3}J_n(y-z)v_i(t-\vep,y)dy-\intl{\R^3}J_n(x-y)v_i(t,y)dy\Big]dz\VS\hfill
=\intl{|x-z|>\eta}H(\vep,x-z)\Big[\intl{\R^3}J_n(y-z)v_i(t-\vep,y)dy-\intl{\R^3}J_n(x-y)v_i(t,y)dy\Big]dz\VS\hfill +
\intl{|x-z|<\eta}H(\vep,x-z)\intl{\R^3}J_n(y-z)\big[v_i(t-\vep,y)- v_i(t,y)\big]dydz\VS\hfill+\intl{|x-z|<\eta}H(\vep,x-z)\intl{\R^3}\big[J_n(y-z)-J_n(x-y)\big]v_i(t,y)dydz =\mbox{${\underset{i=1}{\overset 3\sum}}$}I_i(\vep,\eta). \ea
$$
 Thanks to the energy inequality, the $L^2$-norm of $v$ is finite for all $t>0$. Hence, for all $n\in\N$,  $\displ \Big[\intl{\R^3}J_n(y-z)v_i(t-\vep,y)dy-\intl{\R^3}J_n(x-y)v_i(t,y)dy\Big]$ belongs to $L^\infty(\R^3)$, uniformly with respect to $\vep$. Therefore it is easy to deduce that, for all $x\in\R^3$  and $\eta>0$,
 $$\lim_{\vep\to0}I_1(\vep,\eta)=0.$$
For the term $I_2$, for $\eta>0$ sufficiently small and $n$ sufficienty large, we get
$$\ba{ll}\displ |I_2(\vep,\eta)|&\displ\leq \dm H(\vep,x)\dm_{L^1(B(x,\eta))}\dm J_n[v_i(t-\vep)-v_i(t)]\dm_\infty\VS&\displ \leq \dm H(\vep,x)\dm_{L^1(B(x,\eta))}\dm v_i(t-\vep)-v_i(t)\dm_{L^\infty(|y|>2M_0R_0)},\ea$$
and, by the continuity proved in Lemma\,\ref{CLI}, we deduce, for all $|x|>2M_0R_0$  and $\eta>0$, $$\lim_{\vep\to0}I_2(\vep,\eta)=0.$$ 
Finally, there holds $$\ba{ll}|I_3(\vep,\eta)|&\displ 
\leq \dm H(\vep,x)\dm_{L^1(B(x,\eta))}\dm J_n[v_i(t)](z)-J_n[v_i(t)](x)\dm_{C(B(x,\eta))}\VS &\displ\leq \dm J_n[v_i(t)](z)-J_n[v_i(t)](x)\dm_{C(B(x,\eta))}, \mbox{ for all }\vep>0.\ea$$
Since, for all $n\in \N$, $J_n[v_i(t)](z)$ is a continuous function of $z$, we deduce that
$$|\lim_{\eta\to0}\lim_{\vep\to0}I_3(\vep,\eta)|\leq 
\lim_{\eta\to0}\dm J_n[v_i(t)](z)-J_n[v_i(t)](x)\dm_{C(B(x,\eta))}
=0.$$
Hence,  for all $n\in\N$,  we have proved that
\be\label{LIMIT}
\lim_{\vep\to0}(v_i(t\!-\!\vep), J_n[H(\vep,x)])= J_n[v_i(t)](x).\ee
\par ii) Trivially, the second term admits limit for $\vep\to0$.
\par iii) For the last term it is enough to note that, for all $n\in\N,\;\nabla J_n[T_i(t-\tau,x)]\in L^1(0,t;L^\infty(\R^3))$ and $v\in L^\infty(0,T;L^2(\R^3))$, then, applying the Lebesgue dominated convergence theorem, we deduce the limit property
\be\label{LIMIIT}\lim_{\vep\to0}\intll s{t-\vep}(v\cdot \nabla J_n[T_i(t-\tau,x)],v)d\tau=\intll s{t}(v\cdot \nabla J_n[T_i(t-\tau,x)],v)d\tau\,.\ee
So that from \rf{PRF}, via \rf{LIMIT}-\rf{LIMIIT}, we have proved 
\be\label{PRFI}\ba{l}\displ J_n[v_i(t)](x)=\mathbb H[J_n[v_i(s)]](t-s,x)
+\intll s{t}(v\cdot \nabla J_n[T_i(t-\tau,x)],v)d\tau,\VS\hskip 5cm\mbox{ for }(t,x)\in (0,T)\times (\R^3\setminus B_{2M_0R_0}).\ea
\ee
Now, we perform the limit as $n\to\infty$. 
We begin by remarking that, for all $t$ and $s$, $\{J_n[v(t)]\}$ and $\{J_n[v(s)]\}$ converge  in $L^2(\R^3)$. So that there exists a subsequence, labeled again by $n$, converging almost everywhere in $(t,x)$ to $v(t,x)$, and to $v(s,x)$ in $L^2(\R^3)$. Therefore, recalling that, for all $t-s>0$, $H(t-s,x-y)\in L^2(\R^3)$,   almost everywhere in $x\in\R^3\setminus B_{2M_0R_0}$ the following limit properties hold:
\be\label{PRFII}\lim_{n}J_n[v_i(t)](x)= v_i(t,x)\mbox{\; and }\lim_{n}\mathbb H[J_n[v_i(s)]](t-s,x)=\mathbb H[v_i(s)](t-s,x)\,.\ee
 Since for $|x|>2M_0R_0$ 
$$\ba{l}\displ\lim_{n}J_n[\nabla T_i(t-\tau,x)](y)=\nabla T_i(t-\tau,x-y)\quad{ and}\quad |\nabla T_i(t-\tau,x-y)|\leq C(x),\VS\hskip 8cm\mbox{ for all }(\tau,y)\in (0,t)\times \ov B_{2M_0R_0},\ea$$ 
and $v\in L^\infty(0,T;L^2(\R^3))$, the following  limit holds for all $|x|>2M_0R_0$:
\be\label{PRFIII}\ba{l}\displ\lim_n\intll st\intl{|y|<2M_0R_0}v(\tau,y)\cdot J_n[\nabla T_i(x,t-\tau)](y)\cdot v(\tau,y)dyd\tau\VS\hskip3cm=\intll st\intl{|y|<2M_0R_0}v(\tau,y)\cdot\nabla T_i(x-y,t-\tau)\cdot v(\tau,y)dyd\tau.\ea\ee
Since $\nabla T(s,z)\in L^1(0,T; L^1(\R^3))$, then $J_n[\nabla T_i(t-\tau,x)](y)$ converges to $\nabla T_i(t-\tau,x-y)$ in $L^1(0,T; L^1(\R^3))$. Moreover, since by \rf{CCTI} $$\dm v(t)\dm_{L^\infty(|y|>2M_0R_0)}\leq c(v_\circ)t^{-\frac12},$$ then the following limit holds for all $|x|>2M_0R_0$:
\be\label{PRFIV}\ba{l}\displ\lim_n\intll st\intl{|y|>2M_0R_0}v(\tau,y)\cdot J_n[\nabla T_i(t-\tau,x)](y)\cdot v(\tau,y)dyd\tau\VS\hskip3cm=\intll st\intl{|y|>2M_0R_0}v(\tau,y)\cdot\nabla T_i(t-\tau,x-y)\cdot v(\tau,y)dyd\tau.\ea\ee
So that passing to the limit in \rf{PRFI}, thanks \rf{PRFII}--\rf{PRFIV}, we get
\be\label{PRFV}\ba{l}\displ v_i(t,x)=\mathbb H[v_i(s)](t-s,x)
+\intll s{t}(v\cdot \nabla T_i(t,x),v)d\tau,\VS\hskip 5cm\mbox{ for }(t,x)\in (0,T)\times (\R^3\setminus B_{2M_0R_0}).\ea\ee
This proves (\ref{RFI}) for $s>0$. Let us show its validity for $s=0$. 
Since for $t>0$ a solution to the heat equation is a continuous function,  and since the weak solution $v$ is a continuous function in $s=0$ in the $L^2$-norm, we easily get  
\be\label{PRFVI}\ba{ll}\displ\lim_{s\to0}\mathbb H[v_i(s)](t-s,x)\hskip-0.2cm&\displ= \lim_{s\to0}\mathbb H[{v_\circ}_i](t-s,x)+\lim_{s\to0}\mathbb H[v_i(s)-{v_\circ}_i](t-s,x)\VSE=\mathbb H[{v_\circ}_i](t,x),\;\mbox{ a.e. in }x\in \R^3.\ea\ee
For the integral term we have to verify that the integral is well posed on $(0,t)\times \R^3$. Noting that 
$$|x|>2M_0R_0\Rightarrow |\nabla T_i(t-\tau,x-y)|\leq c(x),\mbox{ for all }(\tau,y)\in (0,t)\times B_{2M_0R_0},$$ 
 we deduce 
$$\intl{|y|<2M_0R_0}|v\cdot\nabla T_i(t,x)\cdot v|dy\leq c(x)\dm v\dm_2^2\leq c\dm v_\circ\dm_2^2,\mbox{ for all }\tau\in (0,t).$$
Moreover, by virtue of \rf{CCTI} we have
$$\dm v(t)\dm_{L^{\frac{3-2\vep}{1-2\vep}}(|y|>2M_0R_0)}^2\leq 
\dm v(t)\dm_2^{2\frac{1-2\vep}{3-2\vep}}\dm v(t)\dm_{L^\infty(|y|>2M_0R_0)}^{\frac{4}{3-2\vep}}\leq c
\dm v_\circ\dm_2^{2\frac{1-2\vep}{3-2\vep}} c(v_\circ)^{\frac{4}{3-2\vep}}t^{-\frac{2}{3-2\vep}} 
.$$
Therefore, applying H\"oder's inequality, we deduce
$$\intl{|y|>2M_0R_0}|v\cdot\nabla T_i(t,x)\cdot v|dy\leq
\dm\nabla T_i(t,x)\dm_{\frac32-\vep}\dm v\dm_{L^{\frac{3-2\vep}{6-4\vep}}(|y|>2M_0R_0)}^2\leq c(t-\tau)^{-\frac{3-4\vep}{3-2\vep}}\tau^{-\frac{2}{3-2\vep}}.$$
The above estimates and the limit property \rf{PRFVI} allow to make the limit as $s\to0$ in \rf{PRFV}, and complete the proof of estimate \eqref{RFI}. \par Finally, since the solution is regular for $t>T_0$ (see \cite{L}),  we could repeat the above argument lines with obvious simplifications, starting from  \eqref{PRF}, and get \eqref{RFIa}.\chiu
\section{\large Spatial behavior of the weak solution: proof of Theorem\,\ref{CR}}
For all $a,\,b<1$, we set
$$A:=\intll01\tau^{-a}(1-\tau)^{-b}d\tau\,.$$ Hence, we get\be\label{K}\intll0t\tau^{-a}(t-\tau)^{-b}d\tau=At^{1-a-b}, \mbox{ for all }t>0.\ee
To prove Theorem\,\ref{CR}, as a first step, we only prove estimate \rf{CRI} on the interval $(0,1)$ and $|x|>R_1:=\frac {14}{3} M_0R_0$. \par Our starting point is the representation formula \rf{RFI}, that we write as follows
\be\label{RFIM} v(t,x)=w(t,x)+ \mathbb T[u\otimes u](t,x)+ \mathbb T[u\otimes w](t,x)+\mathbb T[w\otimes u](t,x)+\mathbb T[w\otimes w](t,x)\,,\ee
where $w$ is the solution of the Stokes problem whose properties are established in Lemma\,\ref{HS}, and $u=v-w$. Recall that,  
thanks to Lemma\,\ref{ELI}, $u$ satisfies estimate  \rf{ELII}.
We introduce the following decomposition:
\be\label{DO}\mathbb T[a\otimes b](t,x):=\mathbb T^{(1)}[a\otimes b](t,x)+\mathbb T^{(2)}[a\otimes b](t,x)\,,\ee where $$\ba{ll}&\displ \mathbb T^{(1)}[a\otimes b](t,x):=\intll0t\intl{|y|<\frac67|x|}a(\tau,y)\otimes b(\tau,y)\cdot \nabla T(t-\tau,x-y)dyd\tau\,,\VSE \mathbb T^{(2)}[a\otimes b](t,x):=\intll0t\intl{|y|>\frac67|x|}
a(\tau,y)\otimes b(\tau,y)\cdot \nabla T(t-\tau,x-y)dyd\tau\,.\ea$$
We estimate the terms on the right-hand side of \rf{RFIM}. Our task is to prove that all the  terms satisfy the bound given in  \rf{CRI}.
\begin{itemize}\item[i)] For the solution $w$  estimate \rf{CRI} follows from Lemma\,\ref{HS}.\item[ii)] Since $|y|<\frac 67 |x|$, if  
$a$ and $b\in L^\infty(0,T; L^2(\R^3))$, employing estimate \rf{TPE}, we easily deduce
$$|\mathbb T^{(1)}(t,x)|\leq c\intll0t
\intl{|y|<\frac67|x|}(|x-y|+(t-\tau)^\frac12)^{-4}|a||b|dyd\tau\leq c |x|^{-3}t^{\frac12}\sup_{(0,t)}\dm a\dm_2\dm b\dm_2.$$
Since $u$ and $w\in L^\infty(0,T;L^2(\R^3))$ and $\alpha\in [1,3)$, the above estimate ensures that $$\ba{l}|\mathbb T^{(1)}[u\otimes u]|+|\mathbb T^{(1)}[u\otimes w](t,x)|+|\mathbb T^{(1)}[w\otimes u](t,x)|+|\mathbb T^{(1)}[w\otimes w](t,x)|\VS\hskip5.5cm\leq c(v_\circ)(1+|x|)^{-\alpha},\,(t,x)\in (0,1)\times \R^3\setminus B_{R_1}\,.\ea$$
\item[iii)] First of all, we note that
$$|\mathbb T^{(2)}[u\otimes w](t,x)+\mathbb T^{(2)}[w\otimes u](t,x)|\leq 2\intll0t \intl{|y|>\frac67|x|} 
(|x-y|+(t-\tau)^\frac12)^{-4}|u||w|dyd\tau\,.$$ Also, since $|y|>\frac 67|x|>4M_0R_0$, then \rf{CCTI}   and \rf{HSI}  imply in particular \be\label{LIU}|u(t,y)|\leq c(v_\circ)t^{-\frac12}.\ee Hence,  
from the above integral inequality and again using \rf{HSI} for $w$, recalling \rf{K}, it follows that 
$$\ba{l}|\mathbb T^{(2)}[u\otimes w](t,x)|+|\mathbb T^{(2)}[w\otimes u](t,x)|+|\mathbb T^{(2)}[w\otimes w](t,x)|\VS\leq 2c(v_\circ)(1+|x|)^{-\alpha}\hskip-0.2cm\intll0t \intl{|y|>\frac67|x|}\hskip-0.2cm 
(|x-y|+(t-\tau)^\frac12)^{-4}\tau^{-\frac12}dyd\tau=c(v_\circ)(1+|x|)^{-\alpha}\,,\VS \hskip 6cm (t,x)\in (0,T)\times \R^3\setminus  B_{R_1}\,.\ea$$
\item[iv)] In this step,  we give    a first estimate for $\mathbb T^{(2)}[u,u](t,x)$ of the kind \rf{CRI}. Subsequently, employing all together  estimates i)-iv), we  prove \rf{CRI} for $\mathbb T^{(2)}[u,u](t,x)$, completing the proof of the theorem for $t\in (0,1)$. 
\par First of all, using the interpolation between $L^q$-spaces, estimate \rf{ELII} and estimate \rf{LIU} give
\be\label{LIUQ} \dm u\dm_{L^{\frac{6-4\vep}{1-2\vep}}(6,|x|)}\!\leq \dm u\dm_{L^2(6,|x|)}^\frac{1-2\vep}{3-2\vep}\dm u\dm_{L^\infty(6,|x|)}^\frac2{3-2\vep}\!\leq c(v_\circ)t^{-\frac12\frac{1+2\vep}{3-2\vep}}|x|^{-\frac12\frac{1-2\vep}{3-2\vep}}, t>0,\ee where, here and in the following, $\vep\in(0,\frac12)$.
Applying H\"older's inequality, then estimates \rf{LIUQ} and \rf{K}, we get
$$\ba{ll}|\mathbb T^{(2)}[u,u](t,x)|\hskip-0.2cm&\displ\leq c\intll0t\dm \nabla T(t,x)\dm_{\frac32-\vep}\dm u\dm_{L^{\frac{6-4\vep}{1-2\vep}}(|y|>\frac67|x|)}^2d\tau \VSE\leq c(v_\circ)\intll0t(t-\tau)^{-\frac{3-4\vep}{3-2\vep}}\tau^{-\frac{1+2\vep}{3-2\vep}}d\tau\leq c(v_\circ)t^{-\frac1{3-2\vep}}|x|^{-\frac{1-2\vep}{3-2\vep}}.\ea$$
\end{itemize}
Now,  from formula \rf{RFIM} and estimates i)-iv) we get
$$|v(t,x)|\leq c(v_\circ)t^{-\frac1{3-2\vep}}|x|^{-\frac{1-2\vep}{3-2\vep}},(t,x)\in (0,1)\times\R^3\setminus B_{R_1}.$$
On the other hand, taking into account \rf{HSI} and that $u=v-w$, then
\be\label{PCRI}|u(t,x)|\leq c(v_\circ)t^{-\frac1{3-2\vep}}|x|^{-\frac{1-2\vep}{3-2\vep}},\;(t,x)\in (0,1)\times\R^3\setminus B_{R_1}\,.\ee 
  Employing estimate \rf{ELII} and \rf{PCRI}, we modify \rf{LIUQ} as follows
\be\label{LIUQI}\ba{ll}\dm u\dm_{L^{\frac{6-4\vep}{1-2\vep}}(|y|>|x|)}\hskip-0.2cm&\leq \dm u\dm_{L^2(6,|x|)}^\frac{1-2\vep}{3-2\vep}\dm u\dm_{L^\infty(|y|>|x|)} ^\frac2{3-2\vep}\VSE\leq c(v_\circ)t^{\frac{1-2\vep}{6-4\vep}-\frac2{(3-2\vep)^2}}|x|^{-\frac{1-2\vep}{6-4\vep}-\frac{2-4\vep}{(3-2\vep)^2}},\; t\in(0,\!1).\ea\ee
We evaluate $\mathbb T^{(2)}[u,u](t,\frac76x)$ via \rf{LIUQI}:
\be\label{LIUQII}\ba{ll}|\mathbb T^{(2)}[u,u]\mbox{$(t,\frac76x)$}|\hskip-0.2cm&\displ\leq c\intll0t\dm \nabla T(t,\mbox{$\frac76$}x)\dm_{\frac32-\vep}\dm u\dm_{L^{\frac{6-4\vep}{1-2\vep}}(|y|>|x|)}^2d\tau \VSE\leq c(v_\circ) |x|^{-\frac{1-2\vep}{3-2\vep}-\frac{4-8\vep}{(3-2\vep)^2}}\intll0t(t-\tau)^{-\frac{3-4\vep}{3-2\vep}}\tau^{\frac{1-2\vep}{3-2\vep}-\frac4{(3-2\vep)^2}}d\tau\VSE\leq c(v_\circ)t^{-\frac{1+2\vep}{(3-2\vep)^2}}|x|^{-\frac{1-2\vep}{3-2\vep}-\frac{4-8\vep}{(3-2\vep)^2}},\;(t,x)\in (0,1)\times\R^3\setminus B_{R_1} .\ea\ee
Now, from formula \rf{RFIM}, estimates i)-iii) and estimate \rf{LIUQII}, via \rf{HSI}, we modify \rf{PCRI} as follows:
\be\label{PCRII}|u(t,\mbox{$\frac76$}x)|\leq c(v_\circ)t^{-\frac{1+2\vep}{(3-2\vep)^2}}|x|^{-\frac{1-2\vep}{3-2\vep}-\frac{4-8\vep}{(3-2\vep)^2}} ,\;(t,x)\in (0,1)\times\R^3\setminus B_{R_1}\,.\ee 
  Employing estimate \rf{ELII} and \rf{PCRII}, we modify \rf{LIUQI} as follows:
\be\label{LIUQIII}\ba{ll}\dm u\dm_{L^{\frac{6-4\vep}{1-2\vep}}(|y|>\frac76|x|)}\hskip-0.2cm&\leq \dm u\dm_{L^2(6,|x|)}^\frac{1-2\vep}{3-2\vep}\dm u\dm_{L^\infty(|y|>\frac76|x|)} ^\frac2{3-2\vep}\VSE\leq c(v_\circ)t^{\frac{1-2\vep}{6-4\vep}-\frac{2+4\vep}{(3-2\vep)^3}}|x|^{-\frac{1-2\vep}{6-4\vep}-\frac{(2-4\vep)}{(3-2\vep)^2}-\frac{8-16\vep}{(3-2\vep)^3}},\; t\in(0,\!1).\ea\ee
By the same arguments we evaluate $\mathbb T^{(2)}[u,u](t,\frac{49}{36}x)$as 
\be\label{LIUQIV}\ba{ll}|\mathbb T^{(2)}[u,u]\mbox{$(t,\frac{49}{36}x)$}|\hskip-0.3cm&\displ\leq c\intll0t\dm T(t,\mbox{$\frac76$}x)\dm_{\frac32-\vep}\dm u\dm_{L^{\frac{6-4\vep}{1-2\vep}}(|y|>\frac76|x|)}^2d\tau \VSE\leq c(v_\circ) |x|^{-\frac{1-2\vep}{3-2\vep}-\frac{4-8\vep}{(3-2\vep)^2}-\frac{16-32\vep}{(3-2\vep)^3}}\!\!\intll0t\!(t\!-\!\tau)^{-\frac{3-4\vep}{3-2\vep}}\tau^{\frac{1-2\vep}{3-2\vep}-\frac{4+8\vep}{(3-2\vep)^3}}d\tau\VSE\leq c(v_\circ)t^\sigma |x|^\gamma ,\;(t,x)\in (0,1)\times\R^3 \setminus B_{R_1}\,,\ea\ee 
with $\sigma:={\frac{5+4\vep^2-20\vep}{(3-2\vep)^3}}$ and $\gamma:={-\frac{1-2\vep}{3-2\vep}-\frac{4-8\vep}{(3-2\vep)^2}-\frac{16-32\vep}{(3-2\vep)^3}}$. Exponent $\sigma$ is nonnegative for all $\vep\in (0,\frac52-\sqrt5]$. We choose $\vep\in (0,\frac52-\sqrt5)$ such that $\gamma=-\frac32$. If $\alpha \leq \frac32$, then  estimate i)-iii) and estimate \rf{LIUQIV} give 
$$|v(t,\mbox{$\frac{49}{36}$}x)|\leq c(v_\circ)|x|^{-\alpha},\mbox{ for all }(t,x)\in (0,1)\times \R^3\setminus B_{R_1}\,.$$
Therefore, for $R_2:=\frac{49}{36}R_1$ and $\alpha\in[1,\frac32]$, 
\be\label{PEFIa}|v(t,x)|\leq c(v_\circ)|x|^{-\alpha}
\mbox{ for all }(t,x)\in (0,1)\times \R^3\setminus B_{R_2}\,.\ee If $\alpha>\frac32$, then, taking into account estimate \rf{PEFIa}, 
$$\ba{ll}|v(t,x)|\hskip-0.2cm&\leq |w(t,x)|+|\mathbb T[v,v](t,x)|\leq |w(t,x)|+|\mathbb T^{(1)}[v,v](t,x)|+|\mathbb T^{(2)}[v,v](t,x)|,\VSE
\mathbb T^{(1)}[v,v](t,x):=\intll0t\intl{|y|<\frac{|x|}2}|\nabla T(t-\tau,x-y)||v(\tau,y)|^2dyd\tau, \VSE
\mathbb T^{(2)}[v,v](t,x):= \intll0t\intl{|y|>\frac{|x|}2}|\nabla T(t-\tau,x-y)||v(\tau,y)|^2dyd\tau,\VSE \hfill \mbox{ for all }(t,x)\in (0,1)\times\R^3\setminus B_{2R_2}\,.\ea$$
The term $\mathbb T^{(1)}[v,v]$ admits the estimate:
$$|\mathbb T^{(1)}[v,v](t,x)|\leq c\dm v_\circ\dm_2^2|x|^{-3},\mbox{ for all }t\in(0,1).$$
Since $|x|>2R_2$, thanks to \rf{LIUQIV} for $\gamma=-\frac32$, the term $\mathbb T^{(2)}[v,v](t,x)$ admits the estimate:
$$\ba{ll}|\mathbb T^{(2)}[v,v](t,x)|\hskip-0.2cm&\displ\leq \intll0t\dm \nabla T(t-\tau,x)\dm_1\dm v(\tau)\dm_{L^\infty(|y|>\frac{|x|}2>R_2)}^2d\tau\leq c(v_\circ)|x|^{-3}\intll0t(t-\tau)^{-\frac12}d\tau\VSE\leq c(v_\circ)|x|^{-3},\mbox{ for all }t\in(0,1].\ea$$
Since $\alpha<3$, the above estimates and \rf{PEFIa},
for  $\alpha\in[1,3)$, prove
\be\label{PEFI}|v(t,x)|\leq c(v_\circ)|x|^{-\alpha}\,,
\mbox{ for all }(t,x)\in (0,1)\times \R^3\setminus B_{2R_2}\,.\ee 
\par We complete the proof of Theorem\,\ref{CR} for $t>1$. 
\par We consider representation formula \rf{RFI} for $s=1$.
Since the previous arguments ensure that $|v(s,x)|\leq c(v_\circ)|x|^{-\alpha}$, for  all $x\in \R^3\setminus B_{R_2}$, then, thanks to Lemma\,\ref{HS}, we easily deduce that
\be\label{MI}|\mathbb H[v_i(1)](t-1,x)|\leq c(v_\circ)|x|^{-\alpha},\mbox{ for all }t>1\mbox { and }x\in\R^3\setminus B_{2R_2}.\ee
So that we evaluate $\mathbb T[v,v](t,x)$, whose decomposition is
$$\ba{ll}\mathbb T[v,v](t,x)\hskip-0.3cm&:=\mathbb T^{(1)}[v,v](t,x)+\mathbb T^{(2)}[v,v](t,x)\VS\mbox{with}&\VS
\mathbb T^{(1)}[v,v](t,x)\hskip-0.3cm &:=\displ\intll1t\intl{|y|<\frac{|x|}2}|\nabla T(t-\tau,x-y)||v(\tau,y)|^2dyd\tau\VS
\mathbb T^{(2)}[v,v](t,x)\hskip-0.3cm &:=\displ\!\!\intll1t\!\!\intl{|y|>\frac{|x|}2}\!\!\!\!|\nabla T(t-\tau,x-y)||v(\tau,y)|^2dyd\tau,\,\mbox{for all } 
(t,x)\!\in (1,T)\!\times\R^3\!\setminus\!B_{2R_2}.\ea$$
We initially consider $\alpha\in [1,2]$. Taking into account \rf{TPE}, 
the term $\mathbb T^{(1)}[v,v]$ admits the estimate:
$$|\mathbb T^{(1)}[v,v](t,x)|\leq c \intll1t(|x|^2+t-\tau)^{-2}
\dm v_\circ\dm_2^2d\tau\leq c(v_\circ)|x|^{-2},t>1$$
If $\alpha \in (2,3)$, then, in particular we get  $v_\circ \in J^\frac32(\R^3)$, Hence, by the results of Lemma\,\ref{DS}, 
\be\label{EDM}\dm v(t)\dm_2\leq c(v_\circ)t^{-\frac14},t>0.\ee
So that, employing \rf{TPE} and \rf{EDM}, we are able to estimate $\mathbb T^{(1)}[v,v](t,x)$  in the following way:
$$\ba{ll}|\mathbb T^{(1)}[v,v](t,x)|\hskip-0.2cm&\displ\leq c \intll1t\dm\nabla T(t-\tau,x)\dm_{L^\infty(|y|<\frac{|x|}2)}\dm v(\tau)\dm_2^2d\tau\VSE
\leq c|x|^{-3}\intll1t(t-\tau)^{-\frac12}\tau^{-\frac12}d\tau=c(v_\circ)|x|^{-3},t>1\,.\ea$$
Therefore, we have proved that 
\be\label{FEI}\mbox{for all }\alpha\in[1,3),\quad|\mathbb T^{(1)}[v,v](t,x)|\leq c(v_\circ)|x|^{-\alpha},\;t>0.\ee
We consider again the decomposition:
$$\mathbb T^{(2)}[v,v](t,x)=\mathbb T^{(2)}[u,u](t,x)+\mathbb T^{(2)}[u\otimes w](t,x)+\mathbb T^{(2)}[w\otimes u](t,x)+\mathbb T^{(2)}[w\otimes w](t,x).$$
Moreover, we recall that the estimate of item iii) holds uniformly with respect to $t$. Hence
\be\label{FEII}\ba{l}\mathbb T^{(2)}[u\otimes w](t,x)|+|\mathbb T^{(2)}[w\otimes u](t,x)|+|\mathbb T^{(2)}[w\otimes w](t,x)|\leq c(v_\circ)|x|^{-\alpha},\VS\hskip5cm\mbox{ for }\alpha\in[1,3),\;t>1\mbox{ and }x\in \R^3\setminus B_{2R_2} .\ea\ee
We consider the $\mathbb T^{(2)}[u\otimes u](t,x)$, for which we argue as in item iv). Taking into account that $|x|>2{R_2}$, estimate \rf{ELII} and estimate \rf{LIU}, employing the interpolation between $L^q$-spaces, give
\be\label{AFEII}\dm u\dm_{L^{\frac{6-4\vep}{1-2\vep}}(|y|>\frac{|x|}2)}\leq \dm u\dm_{L^2(2,|x|)}^\frac{1-2\vep}{3-2\vep}\dm u\dm_{L^\infty(|y|>\frac23|x|)} ^\frac2{3-2\vep}\leq c(v_\circ)t^{-\frac12\frac{1+2\vep}{3-2\vep}}|x|^{-\frac12\frac{1-2\vep}{3-2\vep}},\;t>0.\ee 
 By using estimate \rf{TPE} for $T(s,z)$ and estimate \rf{AFEII} for $u$, applying H\"older's inequality, we get 
$$\ba{ll}|\mathbb T^{(2)}[u\otimes u](t-\tau,x)|\hskip-0.2cm&\displ\leq c\intll1t
\dm \nabla T(t-\tau,x)\dm_{\frac32-\vep}\dm \dm u(\tau)\dm_{L^{\frac{6-4\vep}{1-2\vep}}(|y|>\frac{|x|}2)}^2d\tau\VSE\leq c(v_\circ)|x|^{-\frac{1-2\vep}{3-2\vep}}\intll1t(t-\tau)^{-\frac{3-4\vep}{3-2\vep}}\tau^{-\frac{1+2\vep}{3-2\vep}}d\tau\leq c(v_\circ)t^{-\frac1{3-2\vep}}|x|^{-\frac{1-2\vep}{3-2\vep}},\VSE \hskip 4cmt>1\mbox{ and }x\in \R^3\setminus B_{2R_2}.\ea$$
The last estimate, together with estimates \rf{FEI} and \rf{FEII}, furnishes
$$|v(t,x)|\leq c(v_\circ)|x|^{-\frac{1-2\vep}{3-2\vep}},t>1,x\in \R^3\setminus B_{2R_2}.$$
Now, for the term $\mathbb T^{(2)}[v\otimes v]$, we employ a bootstrap argument to realize the exponent $\alpha$ of spatial decay.
The last estimate and  estimate \rf{LIOI} give
\be\label{BFEII}|v(t,x)|^2\leq c(v_\circ)|x|^{-\frac43\frac{1-2\vep}{3-2\vep}}\dm v(t)\dm_\infty^\frac23\leq c(v_\circ)|x|^{-\frac43\frac{1-2\vep}{3-2\vep}}t^{-\frac12},t>0,x\in \R^3\setminus B_{2R_2}.\ee
Hence,  recalling that $\dm \nabla T(t-\tau,x)\dm_1\leq c(t-\tau)^{-\frac12}$ and \rf{K}, we get
$$|\mathbb T^{(2)}[v\otimes v]|\leq c(v_\circ)|x|^{-\frac43\frac{1-2\vep}{3-2\vep}}\!\intll1t\!
\dm \nabla T(t-\tau,x)\dm_1\tau^{-\frac12}d\tau\leq c(v_\circ)|x|^{-\frac43\frac{1-2\vep}{3-2\vep}},\;t>1,\;x\in\R^3\setminus B_{2R_2}. $$
The last estimate, together with estimates \rf{FEI},\rf{FEII}, furnishes
$$|v(t,x)|\leq c(v_\circ)|x|^{-\frac43\frac{1-2\vep}{3-2\vep}},t>1,x\in \R^3\setminus B_{2R_2}.$$
 Thanks to the last estimate, we modify \rf{BFEII} as 
\be\label{CFEII}|v(t,x)|^2\!\leq c(v_\circ)|x|^{-(\frac43)^2\frac{1-2\vep}{3-2\vep}}\dm v(t)\dm_\infty^\frac23\!\leq c(v_\circ)|x|^{-\left(\frac43\right)^2\frac{1-2\vep}{3-2\vep}}t^{-\frac12},t>0,x\in \R^3\setminus B_{2R_2}.\ee
If we compare \rf{BFEII} and \rf{CFEII}, then, we find 
that the exponent of spatial decay is increased of a factor $\frac43$. Since this can be made in sequence, after a finite number of steps we arrive to an exponent greater or equal than $\alpha$, proving the final estimate \rf{CRI}. 
The theorem is completely proved.\chiu
\section{Asymptotic time behavior: proof of Corollary\,\ref{CCR}}
Thanks to estimate \eqref{CRI}, to achieve the proof of Corollary\,\ref{CCR} we can limit ourselves to prove estimate \rf{CCRII}. Further, it is enough to prove \rf{CCRII} for $\beta=\alpha$. The instant $T_0\leq c\dm v_\circ\dm_2^4$, given in Corollary\,\ref{CCR}, is the same of Lemma\,\ref{DS} and it is due to Leray in \cite{L}. Moreover, thanks to estimate \rf{DE}, we can prove \rf{CCRII} for $\alpha\in(\frac32,3)$. To this end we start from formula \rf{RFIa} written for $t>2T_0$:
\be\label{PRFVT}\ba{l}\displ v_i(t,x)=\mathbb H[v_i(T_0)](t-T_0,x)
+\intll {T_0}{t}(v\cdot \nabla T_i(t,x),v)d\tau,\;\mbox{ for }(t,x)\in (0,T)\times \R^3.\ea\ee
From Theorem\,\ref{CR} and Lemma\,\ref{DS} we get
\be\label{LIn}|v(T_0,x)|\leq c(v_\circ)(1+|x|)^{-\alpha},\mbox{ for all }x\in \R^3.\ee Then, thanks to the proprieties of the solutions to the Stokes Cauchy problem (see e.g. \cite{K}), we have $|\mathbb H[v(T_0)](t-T_0,x)|\leq c(v_\circ)t^{-\frac\alpha2}$. Thus, to achieve the result we only need to estimate the nonlinear term in \rf{PRFVT}. We set
$$\intll {T_0}{t}(v\cdot \nabla T_i(t,x),v)d\tau=\intll {T_0}{\frac t2}(v\cdot \nabla T_i(t,x),v)d\tau+\intll{\frac t2}{t}(v\cdot \nabla T_i(t,x),v)d\tau=:I_1+I_2.$$
By virtue of \rf{TPE} and \rf{LIn}, for all $\alpha\in(\frac32,3)$, we easily deduce the estimate:
$$|I_1|\leq c(v_\circ)\intll {T_0}{\frac t2}\intl{\R^3}(|x-y|+(t-\tau)^\frac12)^{-4}(1+|y|)^{-2\alpha}dy\leq c(v_\circ)t^{-\frac\alpha2}.$$
Instead, for the term $I_2$ we achieve the result in two steps. For $\alpha\in(\frac32,2]$, recalling that \rf{DE} holds,  we get
$$\ba{l}\displ|I_2|\leq c(v_\circ)\intll{\frac t2}t\intl{\R^3}(|x-y|+(t-\tau)^\frac12)^{-4}(1+|y|)^{(-2+\frac23\alpha)\alpha}\tau^{-\frac\alpha2}dyd\tau\VS\hskip3cm\leq c(v_\circ)t^{-\frac\alpha2}\intl{\R^3}|x-y|^{-2}(1+|y|)^{-\frac23\alpha}dy\leq c(v_\circ)t^{-\frac\alpha2}.\ea$$ Hence, by the first step for $I_2$ and the result for $I_1$ and for the linear part, we conclude
$$|v(t,x)|\leq c(v_\circ)t^{-\frac\alpha2},\mbox{ for all }\alpha\in[1,2].$$  For $\alpha\in(2,3)$,  we invoke this last estimate. Hence, the estimate for $I_2$ becomes:
$$\ba{l}\displ|I_2|\leq c(v_\circ)\intll{\frac t2}t\intl{\R^3}(|x-y|+(t-\tau)^\frac12)^{-4}(1+|y|)^{(-2+\frac\alpha2)\alpha}\tau^{-\frac\alpha2}dyd\tau\VS\hskip3cm\leq c(v_\circ)t^{-\frac\alpha2}\intl{\R^3}|x-y|^{-2}(1+|y|)^{-\frac\alpha2}dy\leq c(v_\circ)t^{-\frac\alpha2}.\ea$$
\par
 Therefore, the proof is completed.
\vskip0.2cm
 {\bf Acknowledgment} -
This research was partly supported by GNFM-INdAM, and by MIUR via the PRIN 2012 {\it ``Nonlinear Hyperbolic Partial Differential Equations, Dispersive and Transport Equations: Theoretical and Applicative Aspects''}.


\begin{thebibliography}{35}

\bibitem{CKN}L. Caffarelli, R.Kohn, L. Nirenberg, {\it Partialregularityofsuitableweaksolutions of the Navier-Stokes equations}, Commun. Pure Appl. Math., {\bf 35} (1982), 771--831 .
 \bibitem{CMI}F. Crispo and P. Maremonti, {\it An interpolation inequality in exterior domains}, Rend. Semin. Mat. Univ. Padova, 112 (2004), 11--39.

 \bibitem{CMSB}F. Crispo and P. Maremonti, {\it On the (x,t) asymptotic properties of solutions of the Navier-Stokes equations in the half-space},  Zap. Nauchn. Sem. POMI, {\bf 318} (2004), 147--202, translation in J. Math. Sci. (N.Y.),  {\bf 136} (2006), 3735--3767.

 \bibitem{CMRWS}F. Crispo and P.Maremonti, {\it A remark on the partial regularity of a suitable weak solution to the Navier-Stokes Cauchy problem}, submitted. 
 
 \bibitem{PD}P. Deuring, {\it Pointwise spatial decay of weak solutions to the Navier-Stokes system in 3D exterior domains}, J. Math Fluid Mech. {\bf 17} (2015), 199--232. 

\bibitem{G} G. P. Galdi, {\it An Introduction to the Mathematical Theory of the Navier-Stokes Equations, Steady-state problems}, Second edition, Springer Monographs in Mathematics, Springer, New York, 2011.

\bibitem{K}G. H. Knightly, {\it On a class of global solutions of the Navier-Stokes equations}, Arch. Rational Mech. Anal., {\bf 21} (1966), 211--245.

\bibitem{Lad}O.A. Ladyzhenskaya, \textit{The mathematical theory of viscous incompressible flow}, 
 Gordon and Breach (1968).
 
\bibitem{L}J. Leray, {\it Sur le mouvement d'un liquide visqueux emplissant l'espace}, Acta
Math., {\bf 63} (1934), 193--248.

\bibitem{MCMP}P. Maremonti, {\it On the asymptotic behavior of the $L^2$-norm of suitable weak solutions to the Navier-Stokes equations in three-dimensional exterior domains}, Comm. Math. Phys., {\bf 198} (1988), 385--400.

\bibitem{MLI}P. Maremonti, {\it A remark on the Stokes problem with initial data in $L^1$}, J. of Math. Fluid Mech., {\bf 13} (2011),  469--480.
\bibitem{MDCDS}P. Maremonti, {\it On the Stokes problem in exterior domains: The maximum modulus theorem}, Discrete Contin. Dyn. Syst.- Ser. A, {\bf 34} (2014), 2135--2171.

\bibitem{scheffer1}V. Scheffer, {\it Hausdorff measure and the Navier-Stokes equations}, Comm. Math. Phys., {\bf 55} (1977),  97--112.

\end{thebibliography}
\end{document}